\documentclass[a4paper,11pt,bold,enumerate,noupper,nocenter,twoside,openright,draft]{report}

\usepackage{psfrag,graphicx}
\usepackage{dcolumn}
\usepackage{amsmath,amssymb}
\usepackage{bm}
\usepackage[dvips]{epsfig,color}
\usepackage[english]{babel}

\usepackage{fancyhdr}

\newcommand{\la}{\langle}
\newcommand{\ra}{\rangle}

\def\be{\begin{equation}}
\def\ee{\end{equation}}
\def\bea{\begin{eqnarray}}
\def\eea{\end{eqnarray}}
\def\sign{\mbox{sgn}}

\def\tr{ \mbox{tr}}
\def\tr{{\rm tr}}
\def\1/2{\frac{1}{2}}

\makeatletter  
\def\cleardoublepage{\clearpage\if@twoside \ifodd\c@page\else% 
\hbox{}% 
\thispagestyle{empty} 
\newpage% 
\if@twocolumn\hbox{}\newpage\fi\fi\fi} 
\makeatother

\title{{\bf Quantum correlations in (1+1)-dimensional systems}\\[0.5cm]
 \large
Universidad de Barcelona.\\
Departamento de Estructura y Constituyentes de la Materia (ECM)}

\author{Enrique Rico Ortega\\
\\
\\
}
\date{\today}

\begin{document}

\maketitle 

\cleardoublepage
\newpage

\pagenumbering{roman}
\setcounter{page}{1}

%\vspace*{\stretch{1}}
%\begin{flushright}

%\end{flushright}
%\vspace*{\stretch{2}}

\vspace*{\stretch{1}}
\begin{center}

{\bf{Tesis presentada para optar al grado de doctor, Mayo 2005.}}

\bigskip

\emph{Programa de doctorado:} {\bf{F\'{\i}sica Avanzada.}} \emph{Bienio:} {\bf{2001-2003.}}

\bigskip

\emph{Thesis advisor:} {\bf{Dr. Jos\'e Ignacio Latorre Sent\'{\i}s.}}

\bigskip

\emph{Tribunal:} {\bf{Dr. A. Ac\'{\i}n, Dr. M. Baig, Dr. J.I. Cirac, Dr. R. Fazio, Dr. A.J. Leggett, Dr. M. Lewenstein, Dr. P. Pascual.}}

\end{center}
\vspace*{\stretch{2}}

\cleardoublepage
\newpage

\vspace*{\stretch{1}}
\begin{flushright}
a las tres Julias.
\end{flushright}
\vspace*{\stretch{2}}

\chapter*{Agradecimientos}
\addcontentsline{toc}{chapter}{Agradecimientos}

Muchas personas han colaborado en este proyecto o me han acompa\~nado y aconsejado durante estos \'ultimos cuatro a\~nos. En la parte m\' as acad\' emica, tengo que destacar a dos personas: Guifr\'e Vidal con quien he disfrutado aprendiendo y trabajando; y Jos\'e Ignacio Latorre, mi jefe, que me gui\'o durante el proyecto y con quien mantuve innumerables discusiones sobre F\'{\i}sica y otros temas m\'as mundanos. 

Especialmente \'utiles fueron las reuniones y los journal clubs que organizamos en la UB. En estos encuentros siempre habia alguna conversaci\'on cr\'{\i}tica con LLuis Masanes, alguna interminable con Rom\'an Or\'us o alg\'un consejo de los seniors Enric Jan\'e y Antonio Ac\'{\i}n. 

Otra de las etapas importantes durante el doctorado fue la estancia que realic\'e en el grupo de A. Ekert y A. Kent, en Cambridge. Alli,  conoc\'{\i} a Alastair, Almut, Angello, Daniel, Jonathan, Lawrence, Marie, Matthias, Roberta, Tom, Toshio y particip\'e de su intensa vida acad\'emica y de uno de los lugares m\'as cosmopolitas que conozco.

Durante estos a\~nos he tenido la oportunidad de trabajar y aprender de forma directa o electr\'onicamente con cient\'{\i}ficos como J.I. Cirac, A. Kitaev, C.A. Lutken, J.K. Pachos, F. Verstraete, J. Vidal, M.M. Wolf; y con el pretexto de interaccionar con otros centros de investigaci\'on he visitado los grupos de N. Cerf y S. Massar, el de H. Briegel, el de T. Leggett, el de R. Laflamme y M. Mosca, el de C.J. Williams con los que disfrut\'e hablando y compartiendo ideas de F\'{\i}sica.

Del aspecto m\'as social, no podr\'e olvidar las incre\'{\i}bles comidas en el fabuloso comedor universitario con mis compa\~neros de departamento: Aleix, Alejandro, Carlos, Dani, Ernesto, Jan, Jaume, Juan, Luca, Miriam, Otger, ToniM, ToniR, Xavi. Sin duda ninguna, agradezco a los amigos que he encontrado en Barcelona: David, Diego (con quienes formamos el tr\'{\i}o resplandor) y Noelia. Con ellos viv\'{\i} situaciones irrepetibles en las Medas y en Panticosa o simplemente viendo alguna pelicula del void en las Blas-Sessions.

Un apoyo fundamental en mi vida en Barcelona ha sido la compa\~n\'{\i}a de Federico y Vicky. Siempre he podido contar con ellos en cualquier situaci\'on y en cualquier instante y gracias a ellos conoc\'{\i} a los situacionistas de Godard y el fetichismo de Truffaut.

Finalmente, reservo las \'ultimas lineas de esta secci\'on a las personas m\'as influyentes e importantes, mi hermana\footnote{Las ilustraciones que aparecen al inicio de cada cap\'{\i}tulo son obra de Julia R. Ortega} y mi madre, a quienes les agradezco toda su confianza y les dedico este trabajo.

\addcontentsline{toc}{chapter}{Contents}
\tableofcontents

%\addcontentsline{toc}{chapter}{List of Figures}
%\listoffigures

\pagenumbering{arabic}
\setcounter{page}{1}

\chapter{Introduction.}

\noindent
\begin{center}
%\resizebox{!}{14.5cm}{\includegraphics{portada/mar.eps}}
\end{center}
\hfill

In the last decades, new ideas and tools coming from quantum information community\cite{Bennett:2000at,Nielsen:2000ne,:2001fz,Esteve:2003uo} have brought new insights and a complementary point of view to problems in quantum physics, in general, and highly correlated quantum systems\cite{Osterloh:2002tx,Osborne:2002hl}, in particular. In fact, it was point by Preskill\cite{Preskill:1999he} that this interdisciplinary area of research could bring  better understanding to questions that appear in many body quantum entanglement and deeper classifications of different phases that appear in strongly entangled systems\cite{Laughlin:1983fy,Wen:2002ez}.

Following these lines, the initial purpose of this work was to study entanglement properties in a many-body quantum system characterized by some local interactions. For this aim, we studied low dimensional magnetic models that can be described by quantum spin chains. Their dynamics are defined by some parameters that, depending on their values, can drastically change the behavior of the ground state and their correlations. In fact, these models, in spite of their simplicity, display a rich structure in the phase diagram, e.g. ordered-disordered magnetic phases or phase transition due to the formation of vortices. Correlations become long range at the critical points and when the temperature of the system is fixed to zero, only pure quantum correlation, i.e. entanglement, can drive the system to the phase transition. So, entanglement is a crucial characteristic in the behavior of quantum systems. We show that the phase diagram of a given system can be described just using their entanglement properties. Due to the complete description of the correlation in one dimensional model, we could also explained the success of some numerical variational method, as density matrix renormalization group, in simulating these systems. The core of this chapter is contained in the published works: \emph{"Entanglement in quantum critical phenomena"} by G. Vidal, J. I. Latorre, E. Rico, A. Kitaev\cite{Vidal:2002rm}, \emph{"Ground state entanglement in quantum spin chains"} by J. I. Latorre, E. Rico, G. Vidal\cite{Latorre:2004eh} and \emph{"Entanlement entropy in the Lipkin-Meshkov-Glick model"} by J. I. Latorre, R. Orus, E. Rico, J. Vidal\cite{Latorre:2004qn}. 

In addition to the first point, special mathematical tools, developed in the area of field theory, can be applied to the study of critical phenomena in one dimensional quantum systems. In our works, we link conformal field theory concepts with entanglement behavior. Conformal field theories have succeeded to describe any property of almost any critical model from the knowledge of general symmetries of the system. As it will be shown, entanglement does not escape from this classification and so, it is also related with the deep ideas of symmetry and universality. Another tool that it is especially important in the description of phase transition is the renormalization group. We proposed how renormalization group flows can be characterized with tools of quantum information theory as majorization and entropy. In fact, it seems that using concepts like majorization, it can be achieved a better understanding in the flows of the hamiltonians or how low energy properties of a system appears as we change the scale (energy, momenta or length) when we prove it. The main part of this chapter is described in: \emph{"Fine-grained entanglement loss along renormalization group flows"} by J.I. Latorre, C.A. Lutken, E. Rico, G. Vidal\cite{Latorre:2004pk}.

The last chapter is focused in the analysis and study of matrix product states which is a formalism to represent the state of any quantum system. This method is specially well suited to  describe in an efficient way one dimensional quantum systems with translational symmetry. In particular, we will address the possibility to realize renormalization group transformation at the quantum level on the state that describes the system, independently of any dynamics. Also, we will show how the fixed point of the transformation can be characterized and we will give a complete classification of quantum states that are fixed point in some relevant cases. The most important aspects of this chapter appears in the article: \emph{"Renormalization group transformations on quantum states"} by F. Verstraete, J.I. Cirac, J.I. Latorre, E. Rico, M.M. Wolf\cite{Verstraete:2004qk}.

\chapter{Scaling of entanglement in (1+1)-dimensional systems.}

\noindent
\begin{center}
%\resizebox{!}{13.3cm}{\includegraphics{portada/escalera.eps}}
\end{center}
\hfill

\section{XXZ and XY models.}

The analysis of entanglement properties in the ground state of two exactly solvable models, XY and XXZ models, is one of the main aims of this work\cite{Vidal:2002rm,Latorre:2004eh}. These models describe quantum spin chains with first neighbor interactions and an applied static magnetic field in the z-direction. Their hamiltonians can be written as,
\be
\begin{split}
&H_{XY}=\sum_{j=1}^N \left( \frac{1+\gamma}{4} \sigma_j^{x} \sigma_{j+1}^{x} + \frac{1-\gamma}{4} \sigma_j^{y} \sigma_{j+1}^{y}- \frac{\lambda}{2}   \sigma_j^{z}  \right)\\
&H_{XXZ}= \sum_{j=1}^N \left(\frac{1}{4}\left( \sigma_j^{x} \sigma_{j+1}^{x} + \sigma_j^{y} \sigma_{j+1}^{y} \right)+ \frac{\gamma}{4}   \sigma_j^{z}  \sigma_{j+1}^{z}  -\frac{\lambda}{2}   \sigma_j^{z} \right)
\end{split}
\ee
where $\{\sigma^{\alpha}_j\}_{\alpha=x,y,z}$ are the Pauli matrices, such that $[\sigma_{j}^{\alpha}, \sigma_{k}^{\beta}]=2i\delta_{j k} \epsilon^{\alpha \beta \theta} \sigma_{j}^{\theta}$. This implies commutation relations in different sites $[\sigma_{j}^{\alpha}, \sigma_{k}^{\beta}]= 0$ with $j \neq k$ but they anticommute at the same site $\{\sigma_{j}^{\alpha}, \sigma_{j}^{\beta}\}=2 \mathbb{I} \delta^{\alpha \beta}$. The parameter $\gamma$ quantifies the anisotropy in the interaction and $\lambda$ the strength of the magnetic field, and we are considering periodic boundary conditions, i.e. $\sigma^{\alpha}_j = \sigma^{\alpha}_{j+N}$

\subsection{Bethe-Ansatz in the XXZ model.}

There are several techniques to determine the low energy behavior of the XXZ model. Recent mathematical tools, like bosonization and mapping to sigma models (see for example \cite{Fradkin:1991ho, Auerbach:1994yp, Gogolin:1998qq} and references therein), have shown the continuum theory that describes this model. Nevertheless, Bethe\cite{Bethe:1931hc}, in 1931, was the first to show how to get the thermodynamical properties using the symmetries that characterize the model and that make it integrable. The hamiltonian of the model can be split as
\be
\begin{split}
H_{XXZ}&= \sum_{j=1}^N \left(\frac{1}{4}\left( \sigma_j^{x} \sigma_{j+1}^{x} + \sigma_j^{y} \sigma_{j+1}^{y} \right)+ \frac{\gamma}{4}   \sigma_j^{z}  \sigma_{j+1}^{z}  -\frac{\lambda}{2}   \sigma_j^{z} \right) \\
&= H_{XX} + \gamma H_{Ising} - \lambda M_{z}
\end{split}
\ee
where $H_{XX}$ corresponds to the interaction in the xy-plane of the spin space and it is the limit of zero anisotropy and magnetic field, i.e. $\{\gamma, \lambda\} \to \{0,0\}$; $H_{Ising}$ describes the interaction in the z-direction of the spin and it is the main contribution in a highly anisotropic model, i.e. $\gamma \gg 1$; finally $M_{z}$ corresponds to the total magnetization in the z-axis. 

In the XXZ model, the hamiltonian, $H(\kappa) = H_{XX} +\kappa H_{\kappa}$, varies as a function of a dimensionless parameter $\kappa$ (the magnetic field or the anisotropy), where $H_{XX}$ and $H_{\kappa}$ commute, so they can be simultaneously diagonalised. While the coupling $\kappa$ is modified, the energy of states changes. When a first excited level decreases its energy enough to get the ground state value, the two levels cross each other (fig.\ref{cross}) and a point of non-analyticity in the ground state is created as a function of $\kappa$. After and before the level-crossing, the ground state is an eigenstate of some properties like the magnetization but this is no longer true at this specific situation. Any point of non-analyticity in the system is identified as a quantum phase transition.
\begin{figure}[!ht]
\begin{center}
%\resizebox{!}{6.0cm}{\includegraphics{grafic/cross.eps}}
\caption[Sequences of level crossing in the XXX model]{\label{cross}Sequences of level crossing in the XXX model, $\gamma =1$, while the magnetic field increases. The initial points in the vertical axes indicates the different spin sectors in which the spectrum of the theory can be organized. The different lines that appear from them indicate the splitting of the degeneracy due to the Zeeman term.}
\end{center}
\end{figure}

As we have just seen, the total spin angular momentum in the z-direction, $M_{z}=\frac{1}{2} \sum_{j=1}^N \sigma_j^{z}$, is a good quantum number to block-diagonalize the interaction. The second symmetry that is explicitly used in the Bethe Ansatz is the translational invariance of the system by any number  of lattice spacing and the periodic boundary conditions.

Due to these symmetries, any translational invariant eigenstate of the total magnetization in the z-direction can be an eigenstate of the XXZ model, so that, the ferromagnetic state, i.e. $|F\ra = | \uparrow \uparrow ... \uparrow \ra$, should be an eigenstate of the hamiltonian. In fact, it can be shown that $H|F\ra =\left( \gamma \frac{N}{4}-\lambda \frac{N}{2} \right) |F\ra =E_F |F\ra$, so it is an energy eigenstate, and $M_{z} |F\ra = \frac{N}{2} |F\ra$, it has maximum magnetization. From the ferromagnetic state, it can be derived a translational invariant state with a well defined magnetization applying the operator $\sigma^-$ that decreases the magnetization in one unit,
\be
|k\ra =\frac{1}{\sqrt{N}} \sum_{l=1}^{N} e^{ik l} \sigma^-_l |F\ra = \frac{1}{\sqrt{N}} \sum_{l=1}^{N} e^{i kl}  |l\ra, \, ~~~~ k=\frac{2 \pi n}{N}, \, ~~ n\in \{1,N\}
\ee 
where $|l\ra$ is the state with the spin at the position $l$ reversed respect to the ferromagnetic state and with energy $\left(E-E_F\right)=-\left(\gamma -\cos{k}\right) + \lambda$. In the case with two reversed spins, the wave function appears as,
\be
|\Psi\ra = \sum_{1\le l_1 < l_2 \le N} a(l_1,l_2) |l_1,l_2 \ra =  \sum_{1\le l_1 < l_2 \le N} a(l_1,l_2) \sigma^-_{l_1} \sigma^-_{l_2}|F\ra
\ee
Bethe's idea was to consider a superposition of plane wave functions: $a(l_1,l_2) = A e^{i(k_1 l_1 +k_2 l_2)} + A' e^{i(k_1 l_2 +k_2 l_1)}$. Imposing that $|\Psi\ra$ is an eigenvector of the hamiltonian, i.e. $H | \Psi \ra=E | \Psi \ra$, then
\be
\begin{split}
2\left[E+\gamma \left(1-\frac{N}{4}\right) \right] &a(l_1,l_2) = a(l_1,l_2+1) + a(l_1-1,l_2),~l_2=l_1+1\\
2\left[E+\gamma \left(2-\frac{N}{4}\right) \right]& a(l_1,l_2) = a(l_1,l_2+1) +a(l_1,l_2-1)+\\
& ~~~~~~~~~+ a(l_1+1,l_2) + a(l_1-1,l_2),~~l_2>l_1+1
\end{split}
\ee
where $E-E_F=2 \lambda - \sum_{j=1,2} \left(\gamma -\cos{k_j}\right)$. Applying the Bethe's ansatz into the last equations, an equivalent relation is derived
\be
2\gamma a(l_1, l_1+1)=a(l_1,l_1)+ a(l_1+1, l_1+1)
\ee
or
\be
\frac{A}{A'}= - \frac{e^{i(k_1+k_2)}+1-2\gamma e^{ik_1}}{e^{i(k_1+k_2)}+1-2\gamma e^{ik_2}}=e^{i\theta}.
\ee
In a general case, where there are $r$ reversed spin respect to the ferromagnetic state, the eigenvector is written as
\be
|\Psi \ra = \sum_{1 \le l_1 < ... < l_r \le N}a(l_1,...,l_r) |l_1,...,l_r\ra, 
\ee 
where the ansatz for the coefficient in the wave function is
\be 
a(l_1,...,l_r)=\sum_{\mathcal{P} \in \mathcal{S}_r} \exp \left(\rm{i}\sum_{j=1}^r k_{\mathcal{P}j} l_j + \frac{\rm{i}}{2}\sum_{i\le j} \theta_{\mathcal{P}i,\mathcal{P}j} \right), 
\ee 
$\mathcal{P}\in \mathcal{S}_r$ denotes one of the $r!$ permutations of $\{ 1,...,r\}$ and $k_i$ and $\theta_{i,j}$ with $(i,j)\in \{ 1,...,r\}$ are the parameters to be determined. Three general conditions hold for these parameters\cite{Orbach:1958mo}:
\be
\label{nonl}
\begin{split}
&\theta_{i,j}=-\theta_{j,i} ~~~~~~~~~~ \forall \{ i,j\}, \\
&\cot\frac{\theta_{i,j}}{2}=\frac{\gamma \sin \frac{k_i-k_j}{2} }{\cos \frac{k_i+k_j}{2}+\gamma \cos \frac{k_i-k_j}{2} } \hspace{3ex} (i,j)\in \{ 1,...,r\}, \\
&N k_i = 2 \pi \mathcal{I}_i + \sum_{j \ne i} \theta_{i,j} ~~~~~~~~~~ i \in \{ 1,...,r\},
\end{split}
\ee
where the integers $\mathcal{I}_i$ are called Bethe quantum numbers and completely determine the state. It is known\cite{Yang:1966ty} that the set $\{ \mathcal{I}_i \}$ with the lowest energy for each $M_z=N/2-r$ satisfies: 
\be 
\mathcal{I}_i =M_z -1 + 2i = \frac{N}{2} - r -1 + 2i \hspace{3ex} i \in \{1,...,r\},
\ee 
and the energy reads
\be 
E=  \frac{\gamma N}{4}- \sum_{i=1}^r (\gamma - \cos k_i ) - \lambda M_z.  
\ee

Therefore, the values $\theta$'s and $k$'s give you the whole description of the ground state of the XXZ model. In this work, we solved the previous non-linear equations system, eq.(\ref{nonl}), with a genetic algorithm, and the maximum numerical error allowed in the parameters $\theta$'s and $k$'s was $10^{-5}$. Also, the spin chains that we considered had an even number of sites $N$, so the eigenvalue of the $M_z$ operator in the ferromagnetic system is the integer $N/2$.

\subsection{Diagonalization and continuum theory in the XY model.}

In this section, we will fix our attention in the XY model,
\be
H_{XY}=\sum_{j=1}^N \left( \frac{1+\gamma}{4} \sigma_j^{x} \sigma_{j+1}^{x} + \frac{1-\gamma}{4} \sigma_j^{y} \sigma_{j+1}^{y}- \frac{\lambda}{2}   \sigma_j^{z}  \right).
\ee
The model without magnetic field was diagonalized in 1961 by Lieb, Schultz and Mattis\cite{Lieb:1961fr}. In 1962, Katsura\cite{Katsura:1962lt} solved it with an applied magnetic field. Pfeuty\cite{Pfeuty:1970dz} solved the Ising limit with a transverse magnetic field in 1970. In 1971, Barouch and McCoy\cite{Barouch:1971lg} got the spin correlation functions. See also the textbooks \cite{Chakrabarti:yx,Christe:cl}.

To diagonalise this hamiltonian several unitary transformations have to be applied. Due to the properties of spin operators that mix commutation and anticommutation relations, it is particularly useful to describe one dimensional spin system by one dimensional spinless fermions, i.e. with anticommuting operators. This step is performed with the Jordan-Wigner transformation that maps the Hilbert space of a spin system into a fermionic one,
\be
\hat{a}_l = \left( \prod_{m<l} \sigma^z_m \right) \frac{\sigma_l^x -i \sigma_l^y}{2}; ~~~~~\{\hat{a}_m,\hat{a}_l^{\dagger}\}=\delta_{ml}, ~~\{\hat{a}_l,\hat{a}_m\}=0 ~~\forall \{l,m\}.
\ee  
Then the hamiltonian is recast into
\be
H_{XY}= \sum_{l=1}^{N}\left(  \frac{\hat{a}_{l+1}^{\dagger} \hat{a}_l + \hat{a}_l^{\dagger} \hat{a}_{l+1} }{2}+ \gamma \frac{\hat{a}_{l+1} \hat{a}_l + \hat{a}^{\dagger}_l \hat{a}^{\dagger}_{l+1} }{2} -\lambda \hat{a}_l^{\dagger} \hat{a}_l   \right).
\ee
The Jordan-Wigner transformation changes the boundary conditions depending on the total z-component of the spin,
\be
\begin{split}
\hat{a}_{N+1}= \left( \prod_{m<N+1} \sigma^z_m \right) \frac{\sigma_1^x -i \sigma_1^y}{2} = i^{-N} e^{i \frac{\pi}{2} \sum_{m=1}^N \sigma_m^z} \hat{a}_1.
\end{split}
\ee
Then, in the sector with $\frac{1}{2}  \sum_{m=1}^N \sigma_m^z =0$, and $\frac{N}{2}$ even (odd), the system has periodic (antiperiodic) boundary conditions.

Due to translational invariance, applying the Fourier transformation almost diagonalizes the hamiltonian,
\be
\hat{d}_k=\sum_{l=1}^N \hat{a}_l e^{-i \frac{2 \pi}{N}kl}\, ~~~~~ \{\hat{d}_k,\hat{d}_p^{\dagger}\} =\delta_{kp}, ~~\{\hat{d}_k,\hat{d}_p\}=0 ~~\forall\{k,p\}, 
\ee
then, the hamiltonian is recast into
\be
\begin{split}
H_{XY}=&\sum_{k=-N/2+1}^{N/2} \left( \cos{\frac{2\pi k}{N}} - \lambda \right) \hat{d}^{\dagger}_k \hat{d}_k+\\
&+ \frac{i\gamma}{2} \sum_{k=-N/2+1}^{N/2} \sin{\frac{2\pi k}{N}} \left( \hat{d}_k \hat{d}_{-k} + \hat{d}^{\dagger}_{k} \hat{d}^{\dagger}_{-k} \right).
\end{split}
\ee

If $\gamma=0$, the theory describes what is called XX model. This model has $U(1)$ invariance which is a continuous symmetry. Coleman-Mermin-Wagner theorem shows that a continuous symmetry can not be spontaneously broken in one dimensional system. Then, there is no transition from a disordered to an ordered phase. Nevertheless, it is still possible to have polynomial decay in the correlation functions via a mechanism described by Berezinskii\cite{Berezinskii:1971cv}, Kosterlitz and Thouless\cite{Kosterlitz:1973lr} with the formation of vortices. In this limit the hamiltonian is completely diagonal,
\be
H_{XX}=\sum_{k=-N/2+1}^{N/2} \left( \cos{\frac{2\pi k}{N}} - \lambda \right) \hat{d}^{\dagger}_k \hat{d}_k,
\ee
The thermodynamical limit is obtained letting $\frac{2\pi k}{N} \to \phi $ and $\frac{1}{N} \sum_{k=-N/2+1}^{N/2} \to \int^{\pi}_{-\pi} \frac{d\phi}{2 \pi}$,
\be
H_{XX}  = \int^{\pi}_{-\pi} \frac{d\phi}{2\pi}  \left( \cos{\phi} - \lambda \right) \hat{d}^{\dagger}_{\phi} \hat{d}_{\phi} =\int^{\pi}_{-\pi} \frac{d\phi}{2\pi} \, \Lambda_{\phi} \, \hat{d}^{\dagger}_{\phi} \hat{d}_{\phi},
\ee
Once the hamiltonian is diagonalized, it is straightforward to obtain the continuum limit. The spectrum $\Lambda_{\phi}$ of the XX model has two Fermi points $\phi_F = \pm \arccos{\lambda}$, i.e. where the energy goes to zero. To develop the low energy theory, we define two fermionic modes, $\left[ \hat{d}_{\phi} \right]_{\phi \to |\phi_F|} = \hat{R}_{\phi}$ and $\left[ \hat{d}_{\phi} \right]_{\phi \to -|\phi_F|} = \hat{L}_{\phi}$, i.e. right and left movers, and expanding the spectrum around the Fermi points, the low energy theory turns out to be
\be
\begin{split}
&H_{XX} \to\\
& v_{F} \int d\phi \, \phi \left( \hat{R}^{\dagger}_{\phi} \hat{R}_{\phi} - \hat{L}^{\dagger}_{\phi} \hat{L}_{\phi}  \right) = -i v_{F} \int dx \,  \left( \hat{R}^{\dagger} \partial_{x} \hat{R} -\hat{L}^{\dagger} \partial_{x} \hat{L} \right)
\end{split}
\ee
with $v_{F}= \left| \frac{\partial \Lambda_{\phi}}{ \partial \phi} \right|_{\phi_F} $ and $x$ the position in the coordinate space. From this expression, the XX model can be seen as a free theory of two noninteracting fermionic fields and it is massless in the interval $\left| \lambda  \right| <1$. For $|\lambda|>1$ the spins in ground state are polarized in the direction of the magnetic field.

For the general case of parameters $(\lambda, \gamma)$, a Bogoliubov transformation recast the theory into a free fermionic one,
\be
\begin{split}
&\hat{b}_k = \cos{\frac{\theta_k}{2}} \hat{d}_k - i \sin{\frac{\theta_k}{2}} \hat{d}^{\dagger}_{-k},\\ 
&\{\hat{b}_p, \hat{b}_k^{\dagger}\}=\delta_{pk}, ~~~ \{\hat{b}_p, \hat{b}_k\}=0 ~~ \forall \{k,p\}, \\
&\cos{\theta_k}= \frac{\cos{\frac{2\pi k}{N}} - \lambda}{\sqrt{\left(\cos{\frac{2\pi k}{N}} - \lambda \right)^2 +\gamma^2 \sin^2{\frac{2\pi k}{N}}}}, 
\end{split}
\ee
then, up to an overall constant, the hamiltonian for the XY model can be rewritten as
\be
\label{diagXY}
H_{XY}=\sum_{k=-N/2+1}^{N/2} \Lambda_{k} \hat{b}^{\dagger}_k \hat{b}_k, ~~~~  \Lambda_{k} = \sqrt{\left(\cos{\frac{2\pi k}{N}} - \lambda \right)^2 +\gamma^2 \sin^2{\frac{2\pi k}{N}}}.
\ee
and in the thermodynamical limit,
\be
H_{XY}=\int_{-\pi}^{\pi} \, \frac{d\phi}{2\pi} \, \Lambda_{\phi} \hat{b}^{\dagger}_{\phi} \hat{b}_{\phi}, ~~~~  \Lambda_{\phi} = \sqrt{\left(\cos{\phi} - \lambda \right)^2 +\gamma^2 \sin^2{\phi}}.
\ee
\begin{figure}[!ht]
\begin{center}
%\resizebox{!}{4.0cm}{\includegraphics{grafic/energy.eps}}
\caption[Energy spectrum $\Lambda_{\phi}$ in the XY model]{Energy spectrum $\Lambda_{\phi}$ for different values of the anisotropy $\gamma$ and magnetic field $\lambda$ in the XY model. The ones with zero energy gap correspond to the critical points. In the figure, critical XX model (red line), critical Ising model (dashed green line), the case [$\gamma=0$, $\lambda=1$] (dot-dashed blue line), the Ising case without magnetic field (straight blue line) and the case [$\gamma=0.5$, $\lambda=0.5$] (dashed violet line).}
\end{center}
\end{figure}

For the continuum limit, we consider first the case $\lambda=1$, i.e. critical XY model. In this case, there is just one Fermi point $\phi_{F}=0$. Expanding the spectrum around this point, the hamiltonian can be written as follows
\be
H_{XY} \to \gamma \int d\phi \, \left| \phi \right| \hat{b}^{\dagger}_{\phi} \hat{b}_{\phi} = \gamma \int_{\phi>0} d\phi \,  \phi  \left( \hat{b}^{\dagger}_{\phi} \hat{b}_{\phi} +   \hat{b}^{\dagger}_{-\phi} \hat{b}_{ -\phi} \right).
\ee
Defining two real fermionic modes, i.e. majorana fermions, such that
\be
\begin{split}
&\check{a}_{R,x}=\int_{\phi>0} \frac{d\phi}{\sqrt{2\pi}}  \left( e^{i\phi x} \hat{b}_{\phi}+e^{-i\phi x} \hat{b}^{\dagger}_{\phi} \right),\\
&\check{a}_{L,x}=\int_{\phi>0} \frac{d\phi}{\sqrt{2\pi}} \left( e^{-i\phi x} \hat{b}_{-\phi}+e^{i\phi x} \hat{b}^{\dagger}_{-\phi} \right),\\
&\{\check{a}_{\alpha,x}, \check{a}_{\beta,y} \} = \delta_{\alpha\beta} ~ \delta(x-y), \, ~~\, \check{a}_{\alpha,x} \check{a}_{\alpha,x} = \frac{1}{2}.
\end{split}
\ee
the system explicitly appears as a massless majorana theory where the anisotropy just redefines its velocity,
\be
H_{XY} \to \frac{-i \gamma}{2}  \int \, dx \, \left( \check{a}_{R,x} \partial_{x} \check{a}_{R,x} -\check{a}_{L,x} \partial_{x} \check{a}_{L,x}  \right).
\ee

For the rest of the parameter space the continuum theory is still recast into a relativistic fermionic theory but with a mass term. Expanding the spectrum $\Lambda_{\phi}$ around the point of minimum energy, that is Fermi points, it is straightforward to see that two different regions have to be described depending on the number of Fermi points,
\be
\phi_F=
\begin{cases}
0 & \lambda +\gamma^2 \geq 1\\
\pm \arccos{\frac{\lambda}{1-\gamma^2}} &\lambda +\gamma^2 < 1
\end{cases}
\ee
For the case $\phi_F =0$ the low energy theory turns out to be
\be
H_{\phi_F =0}\to \int d\phi~ \, \sqrt{m^2+ v_{F}^2 \phi^2} ~\, b^{\dagger}_{\phi} b_{\phi},
\ee
where $m=1-\lambda$ and $v_{F}=\lambda + \gamma^2 -1$. The system is a massive theory with a Klein-Gordon dispersion relation. The mass term describes how much energy has to be given to the system to create an excitation. The critical points are those where the mass goes to zero, in this case $\lambda \to 1$. In the case $\phi_F =\pm \arccos{\frac{\lambda}{1-\gamma^2}} $, two fermionic fields are defined, left and right movers, around the two different Fermi points, $\left[ \hat{d}_{\phi} \right]_{\phi \to |\phi_F|} = \hat{R}_{\phi}$ and $\left[ \hat{d}_{\phi} \right]_{\phi \to -|\phi_F|} = \hat{L}_{\phi}$. Then the low energy hamiltonian is recast into,
\be
\begin{split}
&H_{|\phi_F| =\arccos{\frac{\lambda}{1-\gamma^2}}}\to\\
& \int d\phi \left( \sqrt{m^2+ v_{F}^2 \left(\phi - \phi_{F}\right)^2}  R^{\dagger}_{\phi} R_{\phi} +  \sqrt{m^2+ v_{F}^2 \left(\phi + \phi_{F}\right)^2} L^{\dagger}_{\phi} L_{\phi} \right)
\end{split}
\ee
with $m=\gamma^2 \left( 1-\frac{\lambda^2}{1-\gamma^2} \right)$ and $v_{F}=1-\gamma^2 - \frac{\lambda^2}{1-\gamma^2}$. In this case, the theory appears as a massive relativistic model of two noninteracting fermionic fields and the critical theory appears when $\gamma \to 0$ where the XX model is recovered.
\begin{figure}[!ht]
\begin{center}
%\resizebox{!}{5.0cm}{\includegraphics{grafic/diagram.eps}}
\caption[Critical region for the XY model]{\label{dia}Critical region for the XY model in the plane of the parameters, anisotropy $\gamma$ and magnetic field $\lambda$.}
\end{center}
\end{figure}

\section{Ground state and reduced density matrix in a free theory.}

When the hamiltonian of a quantum system is diagonalized, it is quite simple to obtain the ground state of the system. From eq. (\ref{diagXY}), the XY model can be seen as the sum of decoupled fermionic modes, with a positive dispersion relation $\Lambda_k$ for every mode. Then, the state that minimizes the energy is the one that has no population in any of the modes or it is annihilated by the $\hat{b}_k$ operator,
\be
\hat{b}_k |g\ra = 0, ~~~~ \hat{b}^{\dagger}_k \hat{b}_k |g\ra = 0, ~~~ \forall \, k\in\left[-\frac{N}{2}+1,\frac{N}{2}\right].
\ee

Due to the fact that the system is made of noninteracting fermionic modes, the ground state is the tensor product of the occupation of every mode,
\be
\begin{split}
&|g\ra \la g| = \prod_{\otimes k\in[-\frac{N}{2}+1,\frac{N}{2}]} \rho_k = \prod_{\otimes k\in(-\frac{N}{2}+1,\frac{N}{2})} \begin{pmatrix} n_{k} & 0 \\ 0 & 1-n_{k} \end{pmatrix}\\
&= \prod_{\otimes k\in[-\frac{N}{2}+1,\frac{N}{2}]} \begin{pmatrix} \la \hat{b}^{\dagger}_k \hat{b}_k  \ra & 0 \\ 0 & \la \hat{b}_k \hat{b}^{\dagger}_k  \ra \end{pmatrix},~ \, ~ n_k=0~~\forall \, k\in\left[-\frac{N}{2}+1,\frac{N}{2}\right].
\end{split}
\ee
In fact, the eigenvalues of the density matrix can be obtained from the correlation matrix of a pair of majorana operators (see \cite{Peschel:2003ms,Peschel:2004rr,Kitaev:2000ou}) defined by,
\be
\begin{split}
\check{a}_{2m} = \frac{a^{\dagger}_m+a_m}{\sqrt{2}},&  ~\check{a}_{2m-1} = \frac{a^{\dagger}_m-a_m}{i\sqrt{2}}; ~~ \{\check{a}_m,\check{a}_n\}= \delta_{nm},  ~\check{a}_n\check{a}_n=\frac{1}{2}\\
&\check{a}_{2n} =\frac{1}{\sqrt{2}}\sum_{k=-\frac{N}{2}+1}^{\frac{N}{2}} \left( f_{nk} \hat{b}^{\dagger}_k + f^*_{nk} \hat{b}_k\right), \\
&\check{a}_{2n-1} =\frac{1}{i\sqrt{2}}\sum_{k=-\frac{N}{2}+1}^{\frac{N}{2}} \left(h_{nk} \hat{b}^{\dagger}_k - h^*_{nk} \hat{b}_k \right),
\end{split}
\ee
where $f_{nk}$ and $h_{nk}$ implement the Fourier and Bogoliubov transformations, relating the majorana fermions in the coordinate space with the complex fermions in the momentum representation and they fulfill,
\be
\begin{split}
&\sum_{k=-\frac{N}{2}+1}^{\frac{N}{2}} f_{nk} f^*_{mk} = \sum_{k=-\frac{N}{2}+1}^{\frac{N}{2}} h_{nk} h^*_{mk} = \delta_{nm}; \\
& \sum_{k=-\frac{N}{2}+1}^{\frac{N}{2}} f_{nk} h^*_{mk} = \sum_{k=-\frac{N}{2}+1}^{\frac{N}{2}} f^*_{nk} h_{mk} =g_{nm};\\
&g_{nm}= \frac{1}{N} \sum_{k=-\frac{N}{2}+1}^{\frac{N}{2}} e^{i2\pi(n-m)k/N} \frac{\cos{\frac{2\pi k}{N}} - \lambda - i \gamma \sin {\frac{2\pi k}{N}}}{\left| \cos{\frac{2\pi k}{N}} - \lambda - i \gamma \sin {\frac{2\pi k}{N}} \right|}.
\end{split}
\ee

Therefore, the $2N\times 2N$ correlation matrix of the majorana operators is written as follows,
\be
\label{corre}
\begin{split}
&\Gamma^{(N)}_{nm}=\la  \begin{pmatrix} \check{a}_{2n} \\  \check{a}_{2n-1} \end{pmatrix}  \begin{pmatrix} \check{a}_{2m}, &  \check{a}_{2m-1} \end{pmatrix} \ra \\
&= \frac{1}{2} \sum_{k,k'=-\frac{N}{2}+1}^{\frac{N}{2}} \begin{pmatrix} f_{nk} & f^*_{nk} \\ \frac{h_{nk}}{i} & -\frac{h^*_{nk}}{i} \end{pmatrix}  \la  \begin{pmatrix} b^{\dagger}_{k} \\  b_{k} \end{pmatrix}  \begin{pmatrix} b_{k'}, &  b^{\dagger}_{k'} \end{pmatrix} \ra \begin{pmatrix} f^*_{mk'} & -\frac{h^*_{mk'}}{i} \\ f_{mk'} & \frac{h_{mk'}}{i} \end{pmatrix} \\
&= \frac{1}{2} \sum_{k,k'=-\frac{N}{2}+1}^{\frac{N}{2}} \begin{pmatrix} f_{nk} & f^*_{nk} \\ \frac{h_{nk}}{i} & -\frac{h^*_{nk}}{i} \end{pmatrix}  \begin{pmatrix} n_k  & 0 \\ 0 & (1-n_k) \end{pmatrix} \delta_{k,k'} \begin{pmatrix} f^*_{mk'} & -\frac{h^*_{mk'}}{i} \\ f_{mk'} & \frac{h_{mk'}}{i} \end{pmatrix} \\
&=\frac{1}{2} \sum_{k=-\frac{N}{2}+1}^{\frac{N}{2}} \begin{pmatrix} f_{nk} & f^*_{nk} \\ \frac{h_{nk}}{i} & -\frac{h^*_{nk}}{i} \end{pmatrix}  \rho_k \begin{pmatrix} f^*_{mk} & -\frac{h^*_{mk}}{i} \\ f_{mk} & \frac{h_{mk}}{i} \end{pmatrix} = \frac{1}{2} \begin{pmatrix} \delta_{nm}& -ig_{nm}\\ i g_{mn}&\delta_{nm}\end{pmatrix}
\end{split}
\ee

\begin{figure}[!ht]
\begin{center}
%\resizebox{!}{6.0cm}{\includegraphics{grafic/ruta.eps}}
\caption[Road-map to get the reduced density matrix in the XY model]{Steps followed in the text to diagonalize and obtain the eigenvalues of the density matrix from the correlation matrix.}
\end{center}
\end{figure}

In addition to this, because the XY model can be described as a free theory, applying the Wick's theorem, the expectation value of any set of operators can be decomposed as the sum of contractions of every pair of operators,
\be
\la \hat{a}^{\dagger}_{1}\hat{a}^{\dagger}_{2}\hat{a}_{3}\hat{a}_{4} \ra = \la \hat{a}^{\dagger}_{1}\hat{a}^{\dagger}_{2} \ra \la \hat{a}_{3}\hat{a}_{4} \ra -\la \hat{a}^{\dagger}_{1}\hat{a}_{3}  \ra \la\hat{a}^{\dagger}_{2} \hat{a}_{4} \ra +\la \hat{a}^{\dagger}_{1}\hat{a}_{4} \ra \la \hat{a}^{\dagger}_{2}  \hat{b}_{3} \ra,
\ee
where $\hat{a}^{\dagger}_n=\sum_k e^{-i2\pi kn/N} \left(  \hat{b}^{\dagger}_k \cos{\theta_k/2} -i  \hat{b}_{-k} \sin{\theta_k/2}  \right)$. This means that the state is fully characterized by the expectation values of the one body and two body correlators, i.e. it is a gaussian state. Obviously, this property is still valid although we trace out every degree of freedom except the ones considered in the correlation function, i.e. if we use the density matrix $\rho_{1234}=\tr_{N-1234} |g\ra \la g|$. In particular, we are interesting in finite subsystems of $L$ contiguous sites out of the $N$ sites of the total chain. Again, according to Wick's theorem, this property holds if the reduced density matrix of $L$ sites is the exponential of a quadratic hamiltonian\cite{Chung:2001by}, i.e.
\be
\rho_L= \frac{e^{-\mathcal{H}}}{Z}, ~ \, ~~~ \mathcal{H}=\sum_{n,m=1}^L \left[ \hat{a}^{\dagger}_n \mathcal{A}_{nm} \hat{a}_m +\frac{1}{2} \left(\hat{a}^{\dagger}_n \mathcal{B}_{nm} \hat{a}^{\dagger}_m + h.c. \right)\right] 
\ee
with $Z$ a normalization factor, $\mathcal{A}$ an hermitian matrix and $\mathcal{B}$ an antisymmetric one. Because $\mathcal{H}$ is a quadratic form, it can be diagonalized in a similar way as in the case of the XY model\cite{Lieb:1961fr}, getting a new set of $L$ fermionic modes, such that,
\be
\begin{split}
\mathcal{H}= \sum^{L/2}_{k=-L/2+1} &\tilde{\Lambda}_k \hat{\tilde{b}}^{\dagger}_k \hat{\tilde{b}}_k, ~\rho_L = \tr_{N-L}  |g\ra \la g|= \prod_{\otimes k} \tilde{\rho}_k, ~\tilde{n}_k = \tr{\left(\tilde{\rho}_k \hat{\tilde{b}}^{\dagger}_k \hat{\tilde{b}}_k \right)},\\
&\{\hat{\tilde{b}}_k, \hat{\tilde{b}}_{k'} \}=\{\hat{\tilde{b}}^{\dagger}_k, \hat{\tilde{b}}^{\dagger}_{k'} \}=0, ~~~\{\hat{\tilde{b}}^{\dagger}_k, \hat{\tilde{b}}_k' \}=\delta_{kk'}.
\end{split}
\ee
The real majorana fermions are written in terms of these new fermionic modes as follows,
\be
\begin{split}
&\tilde{a}_{2n} =\frac{1}{\sqrt{2}}\sum_{k=-\frac{L}{2}+1}^{\frac{L}{2}} \left( \tilde{f}_{nk} \hat{\tilde{b}}^{\dagger}_k + \tilde{f}^*_{nk} \hat{\tilde{b}}_k\right), \\
&\tilde{a}_{2n-1} =\frac{1}{i\sqrt{2}}\sum_{k=-\frac{L}{2}+1}^{\frac{L}{2}} \left(\tilde{h}_{nk} \hat{\tilde{b}}^{\dagger}_k - \tilde{h}^*_{nk} \hat{\tilde{b}}_k \right)
\end{split}
\ee
and the $2L\times 2L$ correlation matrix of these majorana fermions are related with the eigenvalues of the density matrix as in the eq.(\ref{corre}),
\be
\label{corL}
\begin{split}
\Gamma^{(L)}_{nm}&=\la  \begin{pmatrix} \tilde{a}_{2n} \\  \tilde{a}_{2n-1} \end{pmatrix}  \begin{pmatrix} \tilde{a}_{2m}, &  \tilde{a}_{2m-1} \end{pmatrix} \ra\\
& =\frac{1}{2} \sum_{k=-\frac{L}{2}+1}^{\frac{L}{2}} \begin{pmatrix} \tilde{f}_{nk} & \tilde{f}^*_{nk} \\ \frac{\tilde{h}_{nk}}{i} & -\frac{\tilde{h}^*_{nk}}{i} \end{pmatrix}  \tilde{\rho}_k \begin{pmatrix} \tilde{f}^*_{mk} & -\frac{\tilde{h}^*_{mk}}{i} \\ \tilde{f}_{mk} & \frac{\tilde{h}_{mk}}{i} \end{pmatrix} \\
&= \frac{1}{2} \begin{pmatrix} \delta_{nm}& -ig_{nm}\\ i g_{mn}&\delta_{nm}\end{pmatrix}
\end{split}
\ee

\section{Analysis of entanglement behavior.}

In the last sections, we have seen how to diagonalize the hamiltonian of two quantum spin models, the XXZ and XY model, with two different methods, the Bethe Ansatz and unitary transformations. These methods allow to obtain the ground state of the system for any value of the parameters (anisotropy $\gamma$ and magnetic field $\lambda$), that define the interaction, and therefore to obtain a complete knowledge of the behavior of the theory. As we have seen, there are some specific values of the couplings that make the theory gapless or in other words, the correlation length diverges. This divergence is not driven by any thermal fluctuation since in our analysis the temperature is always zero and plays no role in these models. Only quantum fluctuations in the ground state can lead the system to the critical region. This fact is the main reason to describe the parameter space of these theories with measures of entanglement or pure quantum correlations that have been developes in the last decades in the area of Quantum Information. For this purpose, given the pure ground state $|g(\gamma, \lambda)\ra$, we perform a bipartition of the system getting the density matrix of L contiguous spins out of the whole chain, i.e. $\rho_{(\gamma, \lambda,L)}= \tr_{N-L}{|g(\gamma, \lambda)\ra \la g(\gamma, \lambda)|}$. Then, we employ the entropy of entanglement\cite{Bennett:1995tk} as a measure of the quantum correlation between both subsystems. This measure is defined by $E_L(|g(\gamma, \lambda)\ra)=S(\rho_{(\gamma, \lambda,L)})=-\tr \rho_{(\gamma, \lambda,L)} \log_2{\rho_{(\gamma, \lambda,L)}}$. Changing the number $L$ of spins in the subsystem we look for the scaling properties of the entropy.

\begin{figure}[!ht]
\begin{center}
%\resizebox{!}{5.0cm}{\includegraphics{grafic/spins.eps}}
\caption[Entropy of entanglement for $L$ contiguous sites]{Entropy of entanglement in a spin chain for a subsystem of $L$ contiguous sites.}
\end{center}
\end{figure}

Entanglement entropy is a positive defined quantity and it is zero for separable state, those one that can be written as the tensor product of its parts. In one dimensional systems of $N$ sites that are translational invariant the entropy of the first $L$ contiguous sites is the same as the entropy of the $N-L$ first sites, i.e. $S_L=S_{N-L}$. A third property that will appear in our analysis is a consequence of the strong subadditivity of the entropy\cite{Lieb:1973yn}: entropy is a concave function respect to the number of sites consider in the bipartition, i.e. $S_L\ge \frac{S_{L-M}+S_{L+M}}{2}$ with $M\in~[0,min\{N-L,L\}]$.

In this section, we will see that just the entropy can describe the complete behavior of the system in the parameter space, what allows us to link the scaling properties with deeper concepts like symmetry and universality. 

\subsection{Entropy of entanglement in the XXZ model}

Although Bethe ansatz gives the eigenvalues of thermodynamical quantities like energy or magnetization in an easy way, it is not so simple to obtain the complete description of the wave function. So that, we were able to compute the ground state for systems up to 18 spins.

In the fig. \ref{size}, we see how the finite size of the system saturates the maximum value of the entropy. The plot shows the curves of the entropy for chains with $N\in\{8,10,12,14,16,18\}$ sites. Although in this case, we are limited with finite chains, the initial behavior of the curve, where the finite size effects have less influence, obeys a logarithm behavior such that
\be 
S_L \sim \frac{1}{3} \log_2{L}.
\ee

\begin{figure}[!ht]
\begin{center}
%resizebox{!}{4.0cm}{\includegraphics{grafic/size.eps}}
\caption[Scaling of entanglement in the isotropic Heisenberg model]{\label{size} Entropy of entanglement for an isotropic Heisenberg chain with no magnetic field applied for several values of the total number of spins. Triangles (N=18), stars (N=16), diamonds (N=14), triangles (N=12), stars (N=10), diamonds (N=8)}
\end{center}
\end{figure}

Also, we show how the entropy is modified as the anisotropy or magnetic field change. In fig. \ref{mag}, we analyze the XXX model (or Heisenberg model) with an applied magnetic field. It is well known that there are two limiting cases in this model: when the field is zero the ground state is a singlet but while the field increases, states with higher magnetization become the ground state. When the magnetic field has a value equal to 2, the ferromagnetic state becomes the ground state which is a separable state with maximum magnetization. In the thermodynamical limit, the interval $|\lambda|<2$ is a critical interval\cite{Bonner:1964mq}, this fact appears in the finite size studying of the entropy with a logarithmic scaling in the interval.

\begin{figure}[!ht]
\begin{center}
%\resizebox{!}{4.0cm}{\includegraphics{grafic/mag18.eps}}
\caption[Entropy of entanglement in the Heisenberg model with an applied magnetic field]{\label{mag} Scaling of the entropy in the Heisenberg model with an applied magnetic field. The plots scale with a logarithmic behavior in the interval $|\lambda|<2$. The upper curve correspond to a zero magnetic field while the lower one is for $\lambda=1.97$; the other curves correspond to $\lambda \in \{0.24,0.68,1.05,1.35,1.59,1.77,1.89\}$. In the limit case $\lambda \to 2$, the entropy vanishes.}
\end{center}
\end{figure}

In the fig. \ref{ani} is represented the behavior of the entanglement entropy in an anisotropic spin chain with $N=18$ sites and no magnetic field applied. It is known\cite{Affleck:1998dv} that for values $|\gamma| \le 1$, the anisotropy remains as a marginal operator, so the long range behavior of the system does not change while the anisotropy is modified. But for $|\gamma| > 1$, this parameter turns to be a relevant deformation and the energy spectrum of the theory develops a gap. Again this behavior is described by the scaling of the entropy, for $|\gamma| \le 1$ the curves overlap with a logarithmic behavior, but for $|\gamma| > 1$ the curves start to bend down and saturate due to the anisotropy.

\begin{figure}[!ht]
\begin{center}
%\resizebox{!}{4.0cm}{\includegraphics{grafic/ani18.eps}}
\caption[Entanglement in the XXZ model]{\label{ani} Entanglement entropy in an anisotropic spin chain with $N=18$ sites. The upper curves that overlap correspond to systems with system with anisotropy $|\gamma| \le 1$, while the three curves that bend down correspond to $\gamma=1.5$,  $\gamma=2.0$ and  $\gamma=2.5$ respectively.}
\end{center}
\end{figure}

\subsection{Entropy of entanglement in the XY model}

In this section we will show the scaling properties of the XY model, for which it is possible to get the thermodynamical limit as we saw in the last section. Several properties make possible a numerical treatment of this model even in the infinite length chain. As we have seen in the last section this model is described by a free fermion theory and the ground state is a gaussian state, only depends on one and two body correlation functions, due to this fact the reduced density matrix of $L$ sites is written in terms of the occupation of $L$ fermionic modes, 
\be
\rho_L=\prod_{\otimes k} \begin{pmatrix} \tilde{n}_k & 0 \\ 0 & 1-\tilde{n}_k \end{pmatrix} =\prod_{\otimes k} \tilde{\rho}_k,
\ee
that it can be obtain from the $2L\times 2L$ correlation matrix of some majorana operators. So that, the numerical cost only grows polynomially with the number of sites in the reduced system and the entropy is just the sum of the entropy for every mode,
\be
S_L= - \sum_k \tr \rho_k \log_2 \rho_k
\ee

The results were obtained using numerical techniques to diagonalize the correlation matrix $\Gamma^{(L)}$, eq.(\ref{corL}). Due to the points just mentioned they coincide with the analytical treatment employed by different authors after our work. Jin and Korepin\cite{Jin:2004ti,Its:2005ji} employed Toeplitz matrix properties to diagonalize the correlation matrix and to get the thermodynamical behavior of the entropy. Peschel\cite{Peschel:bp} got the same results using corner transfer techniques to solve the problem. Keating and Mezzadri\cite{Keating:2004lr} applying random matrix theory also obtained the same conclusions.

One of the main results in this work is the complete description of the behavior of a theory by the scaling properties of the entropy of entanglement. The fig. \ref{plane} shows this fact for the parameter space of the XY model. Points where the entropy diverges correspond to critical regions while the normal phase is characterized by a finite entropy of entanglement. Note and compare the general structure of this plot with the fig. \ref{dia}. In principle, this figure appears quite qualitative but scanning through the different regions we will realize about the amount of information in this plot. 

\begin{figure}[!ht]
\begin{center}
%\resizebox{!}{8.0cm}{\includegraphics{grafic/ffig3.eps}}
\caption[XY phases characterized by entropy of entanglement in the plane of magnetic field and anisotropy]{\label{plane} Entropy of entanglement in a XY spin chain for $L=100$ contiguous sites as a function of the anisotropy $\gamma$ and magnetic field $\lambda$. The regions where the entropy diverges correspond to the critical point of the theory.}
\end{center}
\end{figure}

The line that corresponds to the XX model, i.e. zero anisotropy ($\gamma =0$), is described by a gapless theory of two noninteracting free fermions. In the fig.\ref{xx}, we see that the entropy as a function of the number of sites $L$ of the subsystem grows unbounded like a logarithm. The leading behavior does not change in the interval of the magnetic field $|\lambda|<1$ but some logarithmic correction are introduced when the field is turned on. In the limit case when $\lambda \to 1$ the entropy vanishes. We have seen in the diagonalization of this model that the magnetic field is a marginal quantity that just redefined the Fermi points of the theory, that is the reason why the entropy is only sensitive to logarithmic corrections in the value of the magnetic field. From the plot \ref{xx}, the entropy is described by the relation,
\be
S \left( L,\gamma=0,\lambda \right)=\frac{1}{3} \log_2 L + \frac{1}{6} \log_2 \left( 1-\lambda^2 \right).
\ee
The error in the coefficients of the fit can be got with usual technique giving a value of the order $10^{-8}$.

\begin{figure}[!ht]
\begin{center}
%\resizebox{!}{5.0cm}{\includegraphics{grafic/xxmod.eps}}
\caption[Scaling entropy for the critical XX model]{\label{xx} Entropy for a reduced system of $L$ spin in the XX model. The upper curve corresponds to the system without magnetic field. In the other curves, it can be seen that while the magnetic field increases, the entropy decreases. At the point of the field $\lambda=1$ the ground state is a ferromagnet and the entropy goes to zero.} 
\end{center}
\end{figure}

There is another critical region characterized by a free majorana fermion, that corresponds to values of the magnetic field $\lambda =1$. This region is described by the upper curve in the fig. \ref{isi} and the fig. \ref{anisi} represents the correction with the anisotropy $\gamma$. So, in this region the entropy grows unbounded. Its increasing and the leading order follows a logarithmic function. When we solve the problem, we saw that the anisotropy redefined the velocity in the theory, so is a marginal correction, and therefore the anisotropy just induced logarithmic corrections. From these plots, the entropy is described by,
\be
S \left( L,\gamma,\lambda=1 \right)=\frac{1}{6} \log_2{L} + \frac{1}{6} \log_2{\gamma},
\ee
where the errors in the fit are of the order $10^{-8}$.

Away from the critical points, the entropy grows like a logarithm until the correlation length defined by the inverse of the gap in the energy spectrum, from that point the entropy is kept constant with the value,
\be
S(L,\lambda \ne 1) = \frac{1}{6} \log_2{\frac{1}{m}},~~~\forall L> 1/m. 
\ee
 
\begin{figure}[!ht]
\begin{center}
%\resizebox{!}{5.0cm}{\includegraphics{grafic/ising.eps}}
\caption[Scaling entropy for the quantum Ising model with a transverse magnetic field]{\label{isi} Entropy for the reduced density matrix of $L$ spins in the Ising line, i.e. anisotropy $\gamma=1$. In this region, there is one fixed point at $\lambda =1$ where the entropy grows unbounded with length of the subsystem. For models with magnetic field $\lambda  \ne 1$, the entropy grows with a logarithm behavior till the correlation length, defined by the gap in the energy spectrum, at that point the curve is saturated by the gap. In the limit case $\lambda=0$ the ground state of the system, if there is no symmetry breaking field, is described by a cat state, that it is the reason for the value one in the entropy.}
\end{center}
\end{figure}

\begin{figure}[!ht]
\begin{center}
%\resizebox{!}{4.0cm}{\includegraphics{grafic/anisot.eps}}
\caption[Anisotropy correction to the scaling entropy in the critical XY region]{\label{anisi} Logarithmic correction to the entropy in the XY critical line due to the anisotropy. The points fit the function $\frac{1}{6} \log_2 {\gamma}$ with errors of the order $10^{-8}$.}
\end{center}
\end{figure}

The plateaus in the entropy away from the critical point are studied in detail in the last two figures in the particular case of the Ising line, when $\gamma=1$ with the variation of the magnetic field. In fig.\ref{magn}, a divergent point at $\lambda=1$ shows the critical Ising point. From this point, there are two different behaviors in the plot. If the magnetic field increases the ground state of the system goes to a ferromagnetic state where every spin is aligned in the field direction and the entropy vanishes. If the magnetic field decreases to zero the theory is two-fold degenerate. If there is no symmetry breaking field, the ground state is the symmetric superposition of both vacuums, a cat state, and the entropy goes to one. But this solution is highly unstable, any small perturbation in the hamiltonian breaks this symmetry and chooses one of the vacuum. This scenario is described by the points in the lower part of fig. \ref{magn} which were obtained by DMRG (Density matrix renormalization group) techniques\cite{White:1992fs} because introducing any symmetry breaking field make the system non integrable.
 
\begin{figure}[!ht]
\begin{center}
%\resizebox{!}{4.0cm}{\includegraphics{grafic/ffig1.eps}}
\caption[Dependence of the entropy with the magnetic field in the Ising model]{\label{magn} Entropy of a reduced density matrix for $L=100$ contiguous sites in an Ising chain with magnetic field $\lambda$. In the left part of the figure, the upper points corresponds to the model without any symmetry breaking field. The lower points shows the entropy for a model that includes an small field that forces the symmetry breaking.}
\end{center}
\end{figure}

In fig. \ref{sca}, it is shown how the entropy scales in the Ising model as the magnetic field approaches its critical value. The plot shows that the entropy scales as,
\be
\Delta S^{Ising} \sim \frac{1}{6} \log_2\left( 1-\lambda^2 \right)
\ee

\begin{figure}[!ht]
\begin{center}
%\resizebox{!}{4.0cm}{\includegraphics{grafic/ffig2.eps}}
\caption[Asymptotic behavior of the entropy with the magnetic field close to the critical point]{\label{sca} Scaling of the entropy in the Ising model as the magnetic field approaches its critical values $\lambda=1$. The asymptotes correspond to $\Delta S^{Ising} \sim \frac{1}{6} \log_2\left( 1-\lambda^2 \right)$.}
\end{center}
\end{figure}

These figures, fig. \ref{magn} and fig. \ref{sca}, will become an important part in the description of entanglement loss and renormalization group flows because as we will see in the next section, the flow followed by the Ising model while the magnetic field increases corresponds to a renormalization group flow.

\section{Connection with Conformal Field Theory}

Up to now, we have seen that the leading behavior of the entropy in a critical spin model is given by: $S_L \sim k \log_2{L}$, where $k$ is a constant that does not depend on every detail of the hamiltonian but it seems to remain fixed for a given critical region. An answer of why the entropy has this behavior is given by the structure of any one dimensional local theory, gapless and with a linear dispersion relation. Conformal field theory, (see \cite{Brezin:1989qv,Di-Francesco:1997kv} and references therein) is the area of research that studies the structure of these kind of theories and its physical consequences. Based in general arguments from conformal theory Holzhey, Larsen and Wilczek\cite{Holzhey:1994we} in 1994 showed that the leading order behavior for the entropy in a bipartite scenario is given by: $S_L \sim \frac{c+\bar{c}}{6} \log_2{L}$, where $\frac{c+\bar{c}}{6}$ is the universal coefficient (called central charge) that we found for the spin chain, $\frac{1}{3}$ in the XX and Heisenberg model and $\frac{1}{6}$ in the XY model. Similar results were obtained by Korepin in 2004\cite{Korepin:cy}. Calabrese and Cardy\cite{Calabrese:2004eu} in 2004 extended the analysis to massive theories and different partition of the system. Also, scaling entropy has been studied in random quantum critical points by Refael and Moore\cite{Refael:2004gd} getting non-rational effective central charges.

Conformal field theory has classified the properties of any critical field theory with linear energy dispersion in one dimension just from general arguments like the symmetry of the order parameter at the phase transition, leading to the concept of universality of the model, i.e. different physical realizations of a system can be described by the same theory. The coefficient $c$ that appears in the constant term of the entropy is an omnipresent quantity in the classification of conformal field theories, it is the central charge or conformal anomaly. The central charge is one of the quantities that defines the universality and properties of a given critical system\cite{Blote:1986sp,Affleck:1986be}. $c=1/2$ corresponds to a free fermion theory and labels the Ising universality class which agrees with our calculations. $c=1$ corresponds to a free boson theory. In fact, the continuum limit of the XX and Heisenberg model can be seen as a Sine-Gordon model, that is the theory of a bosonic field.

Another remarkable meaning of the central charge is that can be seen as measure of the number of species of a given field in the critical region, i.e it is additive with the number of fields in a theory. So, a theory with two free fermionic fields has total central charge equal to one, this is the reason why $c$ in the XX model is twice the one of the Ising case.  This meaning links with the fact that the central charge appears explicitly in the equation for the entropy that is a measure of the number of degrees of freedom that describe the system.

\section{Scaling of entanglement in the Lipkin-Meshkov-Glick model}

In this section, we will describe\cite{Latorre:2004qn} a model which is placed in a completed connected graph, that is, a model in which every site interacts with each other. This system allows us to study the role of the connectivity in the entanglement properties. For this purpose, we will study a spin system characterized by the Lipkin-Meshkov-Glick model\cite{Lipkin:1965te} (LMG)
\be
H=-\frac{1}{N}\sum_{i<j} \left( \sigma_i^{x} \sigma_j^{x} + \gamma \sigma_i^{y} \sigma_j^{y} \right) - h \sum_i  \sigma_i^{z} 
\ee
where $\sigma_j^{\alpha}$ is the usual Pauli matrix at the $j$ position in the $\alpha$ direction, $N$ is the total number of spins in the systems. This hamiltonian describes a ferromagnetic model with an anisotropy in the $xy$ plane given by the parameter $\gamma$ and a magnetic field $h$ in the $z$ direction. The LMG hamiltonian can be rewritten in terms of the total spin operators $S^{\alpha}=\frac{1}{2} \sum_j \sigma_j^{\alpha}$,
\be
H=\frac{1+\gamma}{N} \left( (S^z)^2+\frac{N}{2}-\vec{S}^2 \right) - 2hS^z + \frac{\gamma - 1}{2N}\left( (S^+)^2+ (S^-)^2\right).
\ee
From this expression, it is straightforward to realize that the ground state of the system lives in the maximum spin sector $S=\frac{N}{2}$. Also, due to the high degree of symmetry of this model, a convenient basis to characterize the system is given by the Dicke states $|\frac{N}{2},m\ra$ that are fully symmetric under any permutation, they are eigenstates of $\vec{S}^2$ with eigenvalue $\frac{N}{2} \left( \frac{N}{2} + 1\right)$ and $S^z$ with eigenvalue $m$.

As we have just mentioned, the ground state of the system for the different values $\gamma$ and $h$ belongs to the permutational symmetric subspace. Therefore, the size of the Hilbert space grows in a polynomial way with the number of sites of the system. In fact, the reduced density matrix of $L$ sites, from the ground state of $N$ spins, is spanned by the set of $L+1$ Dicke states, then the entropy of entanglement is always saturated by $S_L \le \log_2{L+1}$. This properties make that this model can be treated efficiently with numerical methods.

In the fig. \ref{fig:Shg500}, we show the general structure of the ground state characterized by the von Neumann entropy of a bipartition of the system. As in the case of the XY model, we will scan and analyze every region of the parameter space in the following.
\begin{figure}[!ht]
\begin{center}
\caption[Entanglement entropy in the Lipkin-Meshkov-Glick model]{ \label{fig:Shg500} Entanglement entropy for N=500 and L=125 as a function of $h$ and $\gamma$.}
\end{center}
\end{figure}

In the isotropic limit $\gamma \to 1$, the LMG model is diagonal in the Dicke basis. The ground state for a magnetic field $h$ is given by the Dicke state $|\frac{N}{2},M\ra$, its energy is given by $E_0(h,\gamma=1)=-\frac{N}{2}+M^2\frac{2}{N}-2hM$ and magnetization in the $z$ direction,
\be
M=
\begin{cases}
I\left[h\frac{N}{2}\right] & 0\le h \le1 \\
\frac{N}{2} & h\ge 1
\end{cases}
\ee
where $I[x]$ is the round value of $x$.

If instead of the magnetization $M$, we use the number $n=M+\frac{N}{2}$ of "up" spins to describe the ground state, i.e. $|\frac{N}{2},M\ra \equiv |N,n\ra$, we can see that any bipartition of the system is quite simple to realize,
\be
|N,n\ra=\sum_{l=0}^N \sqrt{p_l} |L,l\ra \otimes |N-L,n-l\ra,
\ee
where the subsystems are blocks of size $L$ and $N-L$ sites and $p_l$ is the hypergeometric distribution,
\be
p_l=\frac{\begin{pmatrix}L \\ l \end{pmatrix} \begin{pmatrix} N-L \\ n-l \end{pmatrix} }{\begin{pmatrix} N \\ n \end{pmatrix}}.
\ee
So, the entropy of any subsystem of $L$ sites is given by
\be
S_{L,N}(h,\gamma)=-\sum_{l=0}^L p_l \log_2{p_l}. 
\ee

In the limit $N \gg L \gg 1$, the distribution $p_l$ approaches a gaussian one with mean value $\bar{l}=n\frac{L}{N}$ and variance $\sigma^2 = \frac{n(N-n)}{N} \frac{N-L}{N} \frac{L}{N}$, and the entropy is recast into,
\be
S_{L,N}(h,\gamma=1) = \begin{cases}
\frac{1}{2} \log_2{\left( L\frac{N-L}{N} \right)} + \frac{1}{2} \log_2{\left(1-h^2\right)} & h\in[0,1) \\
0 & h \ge 1
\end{cases}
\ee
where in the interval $h\in[0,1)$ we keep the subleading term $\frac{N-L}{N}$ to make explicit the symmetry $S_{L,N} = S_{N-L,N}$ and in the region $h \ge 1$ the ground state is fully polarized in the direction of the magnetic field, so it is separable and its entropy is zero.

In the anisotropic case, i.e. $\gamma \ne  1$, the LMG hamiltonian is not diagonal but still can be expressed in terms of the Dicke basis, and as we have seen, the cost to get the coefficient of the ground state is just polynomial with the number of sites.  

\begin{figure}[!ht]
\begin{center}
\caption[Entanglement entropy in the LMG model in the anisotropic limit]{\label{fig:Shg0N}Entanglement entropy at $\gamma = 0$ as a function of $h$ for different values of $N$ and $L$. Outside of the critical region, the entropy only depends on the ratio $\frac{L}{N}$.}  
\end{center}
\end{figure}

In the fig. \ref{fig:Shg0N} and \ref{fig:Shg02000}, we plot the dependence of the entropy with magnetic field. Several results appear in these plots that are worth to point out. From the first figure, it can be seen that the value of the entropy for a given value of $h$ and $\gamma$ only depends on the ratio $\frac{L}{N}$. In the limit $h \gg 1$ the ground state of the system is totally polarized in the direction of the magnetic field, and then, the entropy goes to zero. When $h \to 0$, the system is in the symmetric superposition of the polarized vacuums in the $\pm x$ direction, i.e., the ground state is described by a "cat" state or GHZ-like\cite{Greenberger:1989nl} state, so the entropy goes to the value one. As the magnetic field approaches the critical value, $h\to 1$, the entropy displays a logarithm divergence
\be
S_{L,N} \sim - \frac{1}{6}\log_2{|1-h|}
\ee

\begin{figure}[!ht]
\begin{center}
\caption[Entanglement entropy in the LMG model as a function of the magnetic field]{\label{fig:Shg02000}Entanglement entropy as a function of $h$ near the critical point for $\gamma=0$. The full line corresponds to the fitting law $S_{L,N} (h,\gamma) \sim -\frac{1}{6} \log_2{|1-h|}$.}
\end{center}
\end{figure}

In the plot \ref{fig:Sh1L2000}, we display the scaling of the entropy with the number of sites $L$ that are considered in the reduced density matrix. The law that fit the curves is given by
\be
S_{L,N}(h=1,\gamma \ne 1) \sim \frac{1}{3} \log_2{\left( L \frac{N-L}{N} \right)}
\ee
with subleading corrections given by the anisotropy parameter $\gamma$. These corrections are studied in the next figure \ref{fig:Sh1g2000}, from which it can be seen that they are logarithm correction given by
\be
S_{L,N}(h=1,\gamma)-S_{L,N}(h=1,\gamma=0) \sim \frac{1}{6} \log_2{\left(1-\gamma \right)}
\ee 

\begin{figure}[!ht]
\begin{center}
\caption[Scaling of entanglement entropy in the LMG model ]{\label{fig:Sh1L2000}Entanglement entropy as a function of $L$ at the critical point for different  $\gamma$ and $N=2000$. The full line corresponds to the fitting law $S_{L,N} (h=1,\gamma \neq 1) \sim \frac{1}{3} \log_2{\frac{L(N-L)}{N}}$.}
\end{center}
\end{figure}

\begin{figure}[!ht]
\begin{center}
\caption[Entanglement entropy in the LMG model as a function of the anisotropy]{\label{fig:Sh1g2000}Entanglement entropy at the critical point $h=1$ as a function of $\gamma$. The full line corresponds to the fitting law $S_{L,N}(h=1,\gamma) - S_{L,N}(h=1,\gamma=0)\sim \frac{1}{6}  \log_2(1-\gamma)$.}
\end{center}
\end{figure}

This detail analysis of the entanglement entropy of the ground state in the LMG model shows an incredible resemblance with the XY model in one dimensions, although the former is placed in a completely connected graph and the latter in a one dimensional lattice. Due to the symmetry properties of the LMG, we have argued that the entropy can only scale with a logarithm behavior, but it seems that a deeper understanding of scaling in the LMG model is still needed.

\section{Conclusions and related works.}

\subsection{Critical and non critical spin chains.}

The core of this work analyses scaling properties of the entropy of entanglement of the ground state in relevant spin systems. With this detailed study, we have been able to describe the phase diagram of these systems. This analysis and the characterization of the different phases can be linked with concepts like symmetry and universality that appear at the critical points. So, the main results of this sections are summarized as follows:
\begin{enumerate}
\item There exists universal scaling law of entanglement in 1+1 dimensional systems in the critical regions.
\item Entanglement behavior is controlled by conformal symmetry.
\item Away from the critical points entanglement is saturated by mass scales.
\end{enumerate}

As a consequence of these three points and understanding how the size of the Hilbert space of a subsystem scales, it is straightforward to understand the success of DMRG (Density Matrix Renormalization Group) or similar variational methods to describe and implement the ground state of one dimensional systems. In this line, a main question is to get an efficient way to simulate quantum systems in higher dimensions\cite{Verstraete:2004qz}.

\subsection{Entanglement in higher dimensions.}

\begin{table}[h]
\caption{Scaling of entanglement entropy of a subsystem with a typical length $R$.}
\begin{center}
\begin{tabular}{|c ||c| c|} \hline
{\bf{Entropy}} $S_R$& Critical region & Non-critical region \\ \hline \hline
$d=1$ dimension & $\log_2{R} $ &  $\log_2{m}$ \\ \hline
$d>1$ dimensions & \multicolumn{2}{c|}{$R^{d-1}$}\\ \hline
\end{tabular}
\end{center}
\end{table}

One of the first motivations to study entanglement properties in quantum systems was to understand better the origin of black hole entropy\cite{Bekenstein:1994bc} and the area law, i.e. in a three dimensional system, the entropy of a region, with a typical length $R$, is expected to scale as $S\propto R^2$, in a $d$-dimensional system, $S\propto R^{d-1}$. In this context, many works have try to link this entropy with the entanglement entropy between the inside and outside mode of the event horizon (see for example \cite{Bombelli:1986rw,Callan:1994py,Kabat:1994vj,Fiola:1994ir,Kabat:1995eq}). Srednicki in 1993\cite{Srednicki:1993im} using a Klein-Gordon model, explained the area law behavior of the entropy of entanglement as a consequence of the local properties of the interactions and the Schimdt decomposition of the ground state of the system between inside and ouside modes of some region, i.e. $\phi=\sum \mu~\phi_{in}~\phi_{out}$. Because the Schimdt parameters are the same for both subsystems, the entropy of entanglement is the same seeing from inside or outside. So, if there is any dependence in some geometric property of the subsystem, it should be the boundary that they share. From  a quantum information perspective, better bounds and analytic studies have been developed for bosonic\cite{Plenio:2004he} and fermionic fields\cite{Wolf:2005ci}, arriving at similar conclusions. 

\subsection{Non-local interactions and dynamical evolution of entanglement.}

The study of entanglement in non-local systems has been used to describe how entanglement appears in quantum gases or spin glasses\cite{Calsamiglia:2005jt}. This kind of systems gives a model to characterize decoherence. Even, possible consequences to adiabatic quantum computation success arise from these analysis\cite{Latorre:2004ac,Orus:2004xl}. There, it has been shown that adiabatic quantum computation uses exponentially many entanglement when solving NP-complete problems, which has direct implications in classical simulation protocols. Also, it has been applied in the context of dynamical evolution of quantum computers with imperfection\cite{Montangero:2003sk}. Finally, it was noted that the dynamical evolution of entanglement is constrained by properties and symmetries like causality\cite{Calabrese:2005in}

\chapter{Entanglement and RG-flows.}

\noindent
\begin{center}
%\resizebox{!}{14.4cm}{\includegraphics{portada/aguavaso.eps}}
\end{center}
\hfill

The renormalization group (RG) is a method designed to describe how the dynamic of some system changes when we change the scale (distance, energy,...) at which we probe it. A theory at a larger scale employs only a finite part of degrees of freedom from theories at smaller scales, as the irrelevant details are integrated. The decoupling of physics at larger scales from the one a smaller scales is the reason why it is possible to do atomic physics without knowing the final structure of the constituent of the nuclei. So, the description of low-energy physics should not need a detailed knowledge of the laws of Nature at very high momenta. The transition of information from the smaller scale theories to larger scale theories is irreversible, since in such transition we integrate out many irrelevant degrees of freedom. The general aim of renormalization group method is to explain how the decoupling between different scales takes place and how the information is transmitted from scale to scale.

A concrete scheme for renormalization was shown by K. Wilson\cite{Wilson:1974yb,Wilson:1974mb}. One of the main objects in his picture was the space of theories $\mathcal{S}$, as the space of all possible hamiltonians for a given field. So, let $\mu_1$, $\mu_2$ be some scales of momenta or energy, with $\mu_2 < \mu_1$. For each theory $\mathcal{H}\in\mathcal{S}$ there is another theory, $\mathcal{R}_{\mu_1 \mu_2} \mathcal{H} \in \mathcal{S}$ which is the large scale theory (or effective theory), at the scale $\mu_2$, for the original one $\mathcal{H}$ at $\mu_1$. Thus, there is a map $\mathcal{R}_{\mu_1 \mu_2}: \mathcal{S} \rightarrow \mathcal{S}$, with $\mathcal{R}_{\mu_1 \mu_2}\mathcal{R}_{\mu_2 \mu_3}=\mathcal{R}_{\mu_1 \mu_3}$ that defined the action of a semigroup usually called renormalization group. 

The aim of this second work is to analyze how correlations between short and long distance degrees of freedom are lost in RG transformations. To get this goal, we employ tools from Quantum Information to measure the degree of entanglement in the vacuum of a theory. Mainly, we will perform a quantitative analysis in quantum Ising model with a transverse magnetic field as the paradigm of a gaussian fermionic system. In the last part, we will describe the flow of the theory in the space of hamiltonians just with quantities like entropy or majorization.

\section{RG transformation of free theories.}

Any renormalization group transformation\cite{Shankar:1994vs,Morris:1994ie,Berges:2000ew,Ball:1994ji} has two steps:
\begin{enumerate}
\item Integration of high energy modes that let a theory that describes the same long-distance properties.
\item Rescaling of parameters that allows to compare the initial with the renormalized theory because at this point they share the same set of degrees of freedom.
\end{enumerate}
The action of the renormalization group is particularly simple in free theories like the XY model. We have seen that the XY model can be seen as a free theory of fermions,
\be
\hat{H}= \int^{1/a}_{-1/a}  d\phi~ \, \Lambda_{\phi} ~\hat{b}^{\dagger}_{\phi} \hat{b}_{\phi} \simeq \int^{1/a}_{-1/a}  d\phi ~ \, \sqrt{ m^2+ v_{F}^2 \phi^2+ \alpha^2 \phi^4} ~\, \hat{b}^{\dagger}_{\phi} \hat{b}_{\phi},
\ee
where we have explicitly regularized the theory with a lattice spacing $a$ or ultraviolet cutoff $1/a$, and in the second equality, we have developed the dispersion relation to order $o(\phi^4)$. Due to the fact that this theory is noninteractive, the high energy modes are decoupled from the low energy ones, i.e. the partition function can be split in two independent factors: $\mathcal{Z}= \lim_{\beta \to \infty} \tr{ e^{-\beta \hat{H}}} = \mathcal{Z}_{|k|<e^{-l}/a} ~ \mathcal{Z}_{|k|>e^{-l}/a}$, where $l$ is the scaling parameter and the hamiltonian for the low energy modes is written as,
\be
\hat{H}_{|\phi|<e^{-l}/a}= \int^{ e^{-l}/a}_{- e^{-l}/a}  d\phi ~ \, \sqrt{ m^2+ v_{F}^2 \phi^2+ \alpha^2 \phi^4} ~\, \hat{b}^{\dagger}_{\phi} \hat{b}_{\phi}.
\ee
At this stage, it can be seen that in these kind of theories all the non-trivial properties in the RG transformation come from the rescaling of the parameters that is implemented as follows,
\be
\phi = e^{-l} \phi',~~~ \hat{H} = e^{-l} \mathcal{R} \hat{H},~~~b_{\phi} = e^{xl} b_{\phi'}.
\ee
The first two equations take into account the rescaling of the momenta and energy as we change the scale we are using, while the second takes into account the rescaling of the field,  where the scaling dimensions $x$ are fixed by the invariance of the kinetic term $v_{F} \phi^2$, in this case $x=\frac{1}{2}$. Finally, the renormalized hamiltonian is recast into,
\be
\mathcal{R} \hat{H}=  \int^{1/a}_{-1/a} d\phi' ~ \, \sqrt{ \left(e^l m\right)^2+ v_{F}^2 \phi'^2 + \left(e^{-l} \alpha\right)^2 \phi'^4} ~\, \hat{b}^{\dagger}_{\phi'} \hat{b}_{\phi'}.
\ee
So, the theory can be seen as the initial one where the parameters of the hamiltonian have been renormalized, i.e. $m \to e^l m$ and $\alpha \to e^{-l} \alpha$. Then, an infinitesimal transformation in the scale, implies
\be
\frac{\partial v_{F}}{\partial l} =0, ~~~\frac{\partial m}{\partial l} = m,~~~\frac{\partial \alpha}{\partial l} = - \alpha.
\ee
These three equations describe how the parameters of the theory change under the renormalization group, they are usually called beta functions. Successive RG transformations define a RG flow in coupling space. In this example, we see the three kinds of behavior in the parameter space of any theory:
\begin{enumerate}
\item Relevant parameter.- An infinitesimal perturbation of the parameter grows under successive RG transformations. So that, the perturbation is important or relevant in the long distance behavior. In our example, it will correspond to the mass term which drives the system from the Ising fixed point towards an infinitely massive theory.
\item Marginal parameter.- A perturbation in the parameter does not change under RG transformations. In our example the kinetic term.
\item Irrelevant parameter.- A perturbation in the parameter will decrease under RG. So, it is irrelevant for the physics at long distance and if we are looking for an effective theory, we can neglect its effect. In our model, this is the case for the parameter $\alpha$.
\end{enumerate}

A fixed point in an RG transformation is defined where the parameters of the theory remain constant. In free theories, the fixed point appears when $\alpha$ and the mass goes to zero, i.e. at the critical points.

\section{Entanglement loss and RG-flows.}

In one dimensional systems, entanglement loss in RG-flows\cite{Latorre:2004pk} can be analyzed in three different steps that we described in what follows. The first step is just a direct consequence of the structure of critical theories of one dimensional systems and so, it is quite general. For the second and third steps, we will fix our attention in massive fermionic theories to develop all these ideas, taking the Ising model as a representative of these kind of models.

\emph{Global loss of entanglement.-} As we mentioned, conformal field theory describes the properties of one dimensional critical systems and looks for its consequences. One of these consequence was shown by Zamolodchikov\cite{Zamolodchikov:1986gt} in 1984, when he stated the C-theorem (see also \cite{Cappelli:1990yc,Forte:1998dx,Forte:1998jy,Anselmi:1999uk}). There, he proves that unitary and Poincare invariant theories that are perturbed with a relevant deformation from a critical point, under RG transformations, they flow to another fixed point. This flow is characterized by a function $C(g_i,\mu)$ called c-function, that depends on the parameters of the theory $g_i$ and the scale at which we test it, $\mu$. This function is monotonically non-increasing, i.e. $-\left( \mu \frac{\partial g_i}{\partial \mu} \right) \frac{\partial }{\partial g_i} C \le 0$  and remains constant at the fixed points of the renormalization group transformations. The fixed value of the function built by Zamolodchikov coincides with the central charge of model at the critical point, i.e. $C(g_i^*,\mu)=c$. Therefore, the central charge in any ultraviolet fixed point, i.e. when all the details, relevant or irrelevant, are still present, is greater or equal than the central charge of the infrared theory or effective theory, i.e. $c^{UV} \ge c^{IR}$. In the last section, we showed that the central charge of a critical point is directly related to the behavior of the entropy of entanglement in the ground state of the theory, $S(L,g_i^*)=\frac{c+\bar{c}}{6} \log_2{L}$. Then, the entanglement entropy in the $UV$ theory is never smaller than the one in the $IR$ theory, i.e. $S^{UV} \ge S^{IR}$. So, the first link between RG-flows and entanglement appears as a consequence of the structure of conformal theories and relates the entropy between two fixed points. This fact tries to explain the C-theorem by an entropic reasoning\cite{Fiola:1994ir,Casini:2004bw} and points at the loss of entanglement as the cause for the irreversibility in the RG flow.

\emph{Monotonic loss of entanglement.-} Looking for a deeper description of the loss of entanglement under RG transformation, at this second step we will follow the RG trajectories in the Ising model and check that this loss appears step by step in the flow. Our analysis comes from the observation (fig. \ref{magn}) that entanglement entropy itself behaves as a c-function in free fermionic theories, so it is a monotonically non-increasing function in the flow of the mass. In fact, we saw that close to the critical point, $\Delta S(L,\lambda)\sim -\frac{1}{6} \log_2{\left(1-\lambda^2\right)}$. From the plot we observe that there are to limit points, if the magnetic field $\lambda >> 1$ then the entropy goes to zero but in the phase where $\lambda \to 0$ the entropy approaches the unit. This value is characteristic of cat states of GHZ states which are a symmetric superposition of two orthogonal and polarized ground states, i.e. $|\psi\ra = \frac{1}{\sqrt{2}}\left( |\uparrow \uparrow \cdots \uparrow \ra + |\downarrow \downarrow \cdots \downarrow \ra \right)$. Nevertheless this state is highly unstable and any small perturbation of the system, makes it fall in one of the orthogonal vacuums. Then, if an small symmetry breaking field is applied in the hamiltonian, the entropy will approach to zero, where the ground state is totally polarized and separable.

\emph{Fine-grained loss of entanglement.-} (see also \cite{Orus:2005jq}) This third step shows that, at least, for fermionic gaussian theories, the loss of entanglement entropy under RG transformations is just the consequence of the fact that initial and final states of the system are linked by majorization relations (see appendix \ref{entan}). As we will sketch, these transformations are deeply connected with unitary concepts which links with the fact that in the C-theorem one of the main assumptions is the unitary of a theory.

In the second chapter, we have seen that the reduced density matrix of the Ising model can be written in terms of normal fermionic modes, i.e. 
\be
\rho= \prod_{\otimes \phi \ge 0} \rho_{\phi}.
\ee
$\rho_{\phi}$ is the reduce density matrix for the $\phi$ mode, that written in terms of the occupation numbers, is recast into
\be
\rho_{\phi}=\frac{1}{Z_{\phi}} \left( |0\ra \la 0 | + e^{-\omega_{\phi}(\lambda)} |1\ra \la 1|\right),
\ee
where $Z_{\phi}$ is a normalization factor, $\omega_{\phi}$ is the dispersion relation for the $\phi$ mode (their explicit expression for the XY and XXZ model was given by Peschel and collaborators\cite{Peschel:1999xy,Peschel:bp}), $\lambda$ is the magnetic field that characterizes the behavior in the Ising model. 

In what follows, we will see which are the conditions in the dispersion relation in order to fulfill majorization relations (see appendix \ref{entan}) mode by mode and at every infinitesimal step of the RG flow.

The density matrix $\tilde{\rho}_{\phi}$ majorizes  $\rho_{\phi}$, written $\rho_{\phi} \prec \tilde{\rho}_{\phi}$, if and only if they are related by a positive, unital and trace-preserving map, i.e. a double stochastic transformation. In terms of the vector of eigenvalues of each density matrix implies that the vector $\Lambda(\rho_{\phi})$ is given by a probabilistic combination of permutations of the vector of eigenvalues $\Lambda(\tilde{\rho}_{\phi})$, i.e.
\be
\Lambda(\rho_{\phi})=\sum_j p^{(j)}_{\phi} P_j \Lambda(\tilde{\rho}_{\phi})= D_{\phi} \Lambda(\tilde{\rho}_{\phi}),
\ee
where $D_{\phi}$ is a double stochastic matrix,  the real numbers $p^{(j)}_{\phi}$ are such that $0 \le p^{(j)}_{\phi} \le 1$, $\sum_j p^{(j)}_{\phi} =1$ and $\{P_j\}$ is a set of permutation matrices. In the case of fermionic modes, the vector of eigenvalues has just two entries and there are two permutation matrices,
\be
P_0=\begin{pmatrix} 1 &0 \\ 0 & 1 \end{pmatrix},~~~ P_1=\begin{pmatrix} 0 &1 \\ 1 & 0 \end{pmatrix},
\ee
then, any two fermionic density matrix are linked by majorization relations iff
\be
\frac{1}{Z_{\phi}} \begin{pmatrix} 1 \\ e^{-\omega_{\phi}} \end{pmatrix} = \frac{1}{\tilde{Z}_{\phi}} \begin{pmatrix} 1 & e^{-\tilde{\omega}_{\phi}} \\ e^{-\tilde{\omega}_{\phi}} & 1 \end{pmatrix} \begin{pmatrix} p^{(0)}_{\phi} \\ p^{(1)}_{\phi} \end{pmatrix},
\ee
these equations are equivalent to
\be
\begin{split}
p^{(0)}_{\phi}=&\frac{1+e^{-\tilde{\omega}_{\phi}} }{1+e^{-\omega_{\phi}} } \frac{1- e^{-\tilde{\omega}_{\phi}} e^{-\omega_{\phi}} }{1-e^{-2\tilde{\omega}_{\phi}}  }, \\
p^{(1)}_{\phi}=&\frac{1+e^{-\tilde{\omega}_{\phi}} }{1+e^{-\omega_{\phi}} } \frac{ e^{-\omega_{\phi}}- e^{-\tilde{\omega}_{\phi}} }{1-e^{-2\tilde{\omega}_{\phi}}  } .
\end{split}
\ee
with the constraints $p^{(0)}_{\phi}+p^{(1)}_{\phi}=1$, $0\le p^{(0)}_{\phi} \le 1$ and $0\le p^{(1)}_{\phi} \le 1$.

Parametrizing any flow in the space of the density matrix by the variable $\tau$, the dispersion relation will be a function of this parameter $\omega_{\phi}=\omega_{\phi}(\tau)$. An infinitesimal transformation that brings $\tau \to \tilde{\tau}=\tau + \Delta \tau$, will change the dispersion relation to  $\omega_{\phi}=\tilde{\omega}_{\phi}-\frac{\partial \tilde{\omega}_{\phi}}{\partial \tau} \Delta \tau + O\left((\Delta \tau)^2 \right)$, where $\tilde{\omega}_{\phi}=\omega_{\phi}(\tilde{\tau})$.

Then, a simple analysis shows that majorization holds mode by mode at any infinitesimal step in the flow of $\tau$ iff 
\be
0\le \frac{\partial \tilde{\omega}_{\phi}}{\partial \tau} \frac{\Delta \tau}{e^{\tilde{\omega}_{\phi}}-e^{-\tilde{\omega}_{\phi}} }  \le 1
\ee
from this equation, it can be seen the three aspect in the flow and the dispersion relation that have to be checked: (\emph{i}) the increasing behavior of the dispersion relation from $\frac{\partial \tilde{\omega}_{\phi}}{\partial \tau}$; (\emph{ii}) the monotonic behavior of the parameter $\tau$; (\emph{iii}) the sign of the dispersion relation.

Usually, the dispersion relation is positive defined, so that, the main constraint for $\omega_{\phi}$ implies that: $\sign{\left(\frac{\partial \tilde{\omega}_{\phi}}{\partial \tau}\right)}=\sign{\left(\Delta \tau\right)}$

Finally, when majorization holds mode by mode, the whole density matrix of the system, also fulfills these relations. This statement is due to the fact that the tensor product of double stochastic matrices defines a new double stochastic matrix $\mathcal{D}$,
\be
\Lambda(\rho) = \prod_{\otimes \phi} \Lambda(\rho_{\phi}) = \prod_{\otimes \phi}D_{\phi} \Lambda(\tilde{\rho}_{\phi})= \left( \prod_{\otimes \phi}D_{\phi} \right) \Lambda(\tilde{\rho}) = \mathcal{D} \Lambda(\tilde{\rho}).
\ee

As an explicit example, we can see that the Ising model fulfills every conditions in the RG flow in order to relate the eigenvalues of the density matrix with majorization relations and so the loss of entanglement is the consequence of this fact. 

We have seen that the RG flow in this model is characterized by the increasing of the mass term, $\Delta m \ge 0$, so this variable parametrizes the flow of the model, i.e. $\tau \equiv m$. We also know that in the Ising case, $m=|1-\lambda|$, where $\lambda$ is the magnetic field. Then, the departure from the critical point can be seen as the change of the magnetic field around $\lambda =1$. Peschel, Kaulke and Legeza\cite{Peschel:1999xy} showed that the dispersion relation in this model is:
\be
\omega_{\phi}(\lambda)=
\begin{cases} 
(2\phi+1) \pi \frac{I\left(\sqrt{1-(1/\lambda)^2} \right)}{I\left(1/\lambda \right)} & \lambda > 1 \\
2 \phi \pi  \frac{I\left(\sqrt{1-(\lambda)^2} \right)}{I\left(\lambda \right)} & \lambda < 1 
\end{cases}
\ee
where $I(x)$ denotes the complete elliptic integral of the first kind\cite{Gradshteyn:1994pe}.
\begin{figure}[!ht]
\begin{center}
%\resizebox{!}{3.5cm}{\includegraphics{grafic/pesch.eps}}
\caption[Dispersion relation in the Ising model]{Dispersion relation $\omega(\lambda)$ for the Ising model. The first plot shows the function when $\lambda<1$ while the second plot is for $\lambda >1$.}
\end{center}
\end{figure}

From the explicit expressions of the dispersion relation for the Ising model, it is quite simple to see that $\frac{\partial \omega_{\phi}(m)}{\partial m} \ge 0$ and so majorization hold for every mode $\phi$. There is just one last comment about the zero mode in the unbroken phase, when $\lambda < 1$. The eigenvalues of this mode are $\Lambda(\rho_0)=\frac{1}{2} \left(1,1 \right)$ independently of the value of the magnetic field. This special behavior produces a cat state in the limit $\lambda \to 0$, i.e. in a local spin basis $|\psi\ra = \frac{1}{\sqrt{2}}\left( |\uparrow \uparrow \cdots \uparrow \ra + |\downarrow \downarrow \cdots \downarrow \ra \right)$. This state is quite unstable to any perturbation in the hamiltonian and in fact it violates the clustering principle. If a symmetry breaking field is applied to the hamiltonian, the state will flow to one of the polarized vacuum. 

\section{Conclusions and outlook.}

We have seen that a deeper analysis using tools from quantum information to describe and characterize pure quantum correlations has revealed a highly ordered structure in the vacuum of the studied models. This structure seems to imply an order relation in the space of states under scale transformations which we use to point at the entanglement loss as the root of the irreversibility in the renormalization group flows. So, two main ideas appears from this study:
\begin{enumerate}
\item Entanglement is non increasing under RG flows.
\item There exists a close relation between majorization and C-functions.
\end{enumerate}
At least two obvious points remain as open questions, the generalization of these links to interactive theories and if majorization can give new insight to higher dimensional systems.

\chapter{Matrix Product States.}

\noindent
\begin{center}
%\resizebox{!}{14.5cm}{\includegraphics{portada/ntelescopios.eps}}
\end{center}
\hfill

In the last decades, one dimensional models have attracted much attention and effort to understand their behavior. Several reasons motivate this effort: many physical systems can be mapped to these models and new collective phenomena appear in low dimensional systems. In this context, a new sets of methods have been developed to simulate in an efficient way these models and characterize their ground state. The methods that we describe in what follow, share as a building block a set of local finite matrices that implement the information of the system. We will show how to build these matrices, their properties and how can be used to compute, in an efficient way, any correlator. At the end, we will see  how to perform scale transformations within these formalism\cite{Verstraete:2004qk}, how to realize an exact real space renormalization group at the level of the ground state of the system and how it can be found the fixed points of the transformation.

\section{Definition and properties.}

To define the matrix product state, we will show two possible ways to built them. The first one is in the spirit of the Density Matrix Renormalization Group (DMRG) method of S. White\cite{White:1992fs} (see also \cite{Schollwoeck:2004pz}), and it is based in the Schmidt decomposition of a quantum state (see appendix \ref{entan}). The second one, the older point of view, comes from the fact that the matrices can be seen as a bond between the neighbor sites in a chain. This one was the idea of the valence bond model of I. Affleck, T. Kennedy, E.H. Lieb and H. Tasaki\cite{Affleck:1987cm, Affleck:1987cy}, generalized by M. Fannes, B. Nachtergaele and R. F. Werner\cite{Fannes:1990ur}. The link between both point of view was initially given by S. Ostlund and S. Rommer\cite{Ostlund:1995kp}. The results from scaling of entanglement allow us to explain the success of these formalism in one dimensional system. Finally G. Vidal\cite{Vidal:2003jd, Vidal:2004ys}, F. Verstraete, J.I. Cirac, J.J. Garcia-Ripoll and D. Porras\cite{Verstraete:2004lj,Verstraete:2004pu} have shown, in different works, how to generalize these methods to nonlocal interactions, different boundary conditions and how to perform dynamical simulations of quantum systems with classical resources.

Following the fig. \ref{schm}, the Schmidt decomposition of a quantum state between the first site and rest of the chain reads,
\be
|\Psi\ra = \sum_{\alpha_1=1}^D \mu_{\alpha_1} |\alpha_1^{(1)}\ra ~  |\alpha_1^{(2...N)}\ra
\ee
where $\{ |\alpha_1^{(1)}\ra\}$ is an orthonormal basis of the first subsystem and $\{|\alpha_1^{(2...N)}\ra\}$ is an orthonormal set for the rest of the chain, the real numbers $\{\mu_{\alpha_1} ~|~0\le \mu_{\alpha_1} \le 1, ~ \sum_{\alpha_1} (\mu_{\alpha_1})^2 =1 \}$ are the Schmidt coefficients. Finally, the integer $D$ is the Schmidt rank in the decomposition that, for easy writing, we fix to the maximum of any bipartition. If we express the state of the first site in some initial local basis, then the state is written,
\be
|\Psi\ra = \sum_{\alpha_1=1}^D  \sum_{s_1=1}^{d} |s_1\ra \mu_{\alpha_1} \la s_1|\alpha_1^{(1)}\ra ~  |\alpha_1^{(2...N)}\ra =  \sum_{\alpha_1=1}^D  \sum_{s_1=1}^{d} |s_1\ra A_{\alpha_1}^{s_1} ~  |\alpha_1^{(2...N)}\ra
\ee
where in this case, $d$ is the dimension of the local space and $A_{\alpha_1}^{s_1} =  \mu_{\alpha_1} \la s_1|\alpha_1^{(1)}\ra$ is the tensor for the first site. 
\begin{figure}[!ht]
\begin{center}
%\resizebox{!}{2.0cm}{\includegraphics{grafic/schm.eps}}
\caption[Schimdt decomposition picture of a matrix product state]{\label{schm} Schematic first step in the construction of the matrix product state using Schimdt decomposition in a chain with $N$ sites.}
\end{center}
\end{figure}

Carrying the process to the second site,
\be
|\Psi\ra = \sum_{\{\alpha\}=1}^D  \sum_{\{s\}=1}^{d} |s_1\ra A_{\alpha_1}^{s_1} ~ |s_2\ra  A_{\alpha_1\alpha_2}^{s_2} |\alpha_2^{(3...N)}\ra,
\ee
where the sums $\{\alpha\}$ and $\{s\}$ are made over the different configurations of $\alpha_1$, $\alpha_2$, $s_1$ and $s_2$ and the matrix  $A_{\alpha_1\alpha_2}^{s_2} =  \mu_{\alpha_2} \la s_2|\alpha_1\alpha_2^{(2)}\ra$. Then for the whole chain, the state is written as follows,
\be
|\Psi\ra=  \sum_{\{\alpha\}=1}^D  \sum_{\{s\}=1}^{d} |s_1\ra A_{\alpha_1}^{s_1} ~ |s_2\ra  A_{\alpha_1\alpha_2}^{s_2} \cdots |s_{N-1}\ra A_{\alpha_{N-2}\alpha_{N-1}}^{s_{N-1}} |s_N\ra A_{\alpha_{N-1}}^{s_N}.
\ee

Looking at the bulk of an infinite chain, the basis in the bipartition for the right subsystem will be written as,
\be
|\alpha_{L-1}\ra= \sum_{\alpha_L =1}^D \sum_{s_L=1}^d |s_L\ra A_{\alpha_{L-1} \alpha_L}^{s_L} |\alpha_L\ra, 
\ee
where $|\alpha_L\ra= \sum_{\alpha_{L+1} =1}^D \sum_{s_{L+1}=1}^d |s_{L+1}\ra A_{\alpha_L \alpha_{L+1}}^{s_{L+1}} |\alpha_{L+1}\ra$ and if $\{|\alpha_{L-1}\ra\}$ is an orthonormal basis then 
\be
\sum_{s_L=1}^d \sum_{\alpha_L=1}^D \left(A_{\alpha'_{L-1} \alpha_L}^{s_L}\right)^* A_{\alpha_{L-1} \alpha_L}^{s_L} =\sum_{s_L=1}^d  \left(A^{s_L}\right)^{\dagger} A^{s_L} =\mathbb{I},
\ee
which defines the condition in the $A^s$ matrices to be a positive, trace preserving map. Within this approach, the matrices $A^{s_L}$ represent the change in the description of a system when it is considered one site more, they act as a transfer matrix. Every time we insert one site more in the subsystem, a new matrix $A^{s_L}$ link it with the rest of the chain and, in that way, a succession of matrices $\{A^{s_L}\}$ is produced. If this succession has a limit, then we could represent the final state of the chain by this set of matrices. 

So, this method is computationally efficient if the Schmidt rank $D$ is bounded. But this situation appears usually in one dimensional system with local interactions. When we perform the bipartition of the system between L sites and the rest of the chain, the Schmidt rank measures the number of degree of freedom needed to describe the subsystem and is always bounded by the entropy of entanglement, i.e. $D \gtrsim 2^{S_L}$. Using the analysis of the scaling of entanglement in one dimensional models, we know that for finite systems of length $N$ the maximum of the entropy goes as $S_L\sim \log_2{N}$ and for infinite gapped systems, it goes like $S_L\sim \log_2{\frac{1}{m}}$. So, in any of these cases the Schmidt rank is bounded by a constant. Even at the critical points, we know that the entropy grows with the size of the subsystem $L$ as $S_L\sim \log_2{L}$, then, although the Schimdt rank is not saturated in this case, its growth is polynomial with the size of the subsystem, $D\sim L$.

In a translationally invariant state, the whole set of matrices can be represented by just one matrix independently of its position, i.e.  $\{A^{s_L}\} \to A^s$. Then, a translationally invariant state can be described by the product of this matrix as follows,
\be
|\Psi\ra=\sum_{\{s\}=1}^d  \tr \left( B \cdot A^{s_1} \cdots A^{s_N} \right) |s_1, \cdots, s_N  \ra
\ee
where the matrix $B$ implements the boundary conditions of the system. In particular, if the state has periodic boundary condition this matrix is the identity, $B=\mathbb{I}$.

\begin{figure}[!ht]
\begin{center}
%\resizebox{!}{2.0cm}{\includegraphics{grafic/aklt.eps}}
\caption[Valence bond picture of a matrix product state]{\label{aklt}Valence bond picture of a matrix product state. The first line with dark blue spots represents the physical state of the spin chain, while the second line represent the implementation of the state with two ancillae systems per site and a maximally entangled state between neighbor sites.}
\end{center}
\end{figure}

The second point of view, for the matrix product state, comes from the identification of the lower indexes of the matrix $A_{\alpha \beta}^{s}$ with two ancillary subsystems used to implement the state of the physical system $s$. So, this matrix  appears as a projector from the Hilbert space $\mathcal{H}^{(a)}$ of the ancillae to the real one $\mathcal{H}^{(phys)}$, i.e.
\be
\begin{split}
A:~& \mathcal{C}^D \otimes \mathcal{C}^D \to \mathcal{C}^d,\\
&|\alpha,\beta\ra \to A|\alpha,\beta\ra = \sum_{s=1}^d  A_{\alpha \beta}^{s} |s\ra.
\end{split}
\ee
In this way, the correlations in the chain (fig.\ref{aklt}) are implemented by maximally entangled states in the ancillae subsystem of neighbor site, i.e. $\sum_{\alpha=1}^D |\alpha\ra |\alpha\ra$, and the final description of the state is,
\be
|\Psi \ra = A \sum_{\alpha=1}^D |\alpha\ra |\alpha\ra \sum_{\beta=1}^D |\beta\ra |\beta\ra = \sum_{\alpha,\beta=1}^D \sum_{s=1}^d |\alpha\ra  A_{\alpha \beta}^{s} |s\ra |\beta\ra.
\ee
This one was the original method due to Affleck, Kennedy, Lieb and Tasaki (AKLT) to built the exact ground state of antiferromagnetic system of a spin one chain using the symmetric subspace of two spin $1/2$ ancillae.

Once the state of the system is built, it is important to know how to get any correlator from it. In the matrix product state formalism, this task is implemented defining a transfer matrix . For instance, to get the normalization in a translational invariant state,
\be
\la \Psi | \Psi \ra =  \sum_{\{s\}=1}^d   \sum_{\{\tilde{s}\}=1}^d  \la s_1 .. s_N| \tr \left( A^{s_1} .. A^{s_N} \right)^*  \tr \left( A^{\tilde{s}_1} .. A^{\tilde{s}_N} \right) |\tilde{s}_1.. \tilde{s}_N  \ra,
\ee
defining the $D^2 \times D^2$ transfer matrix $E=\sum_{s=1}^d  (A^{s})^* \otimes A^{s}$, the normalization is just $\la \Psi | \Psi \ra =\tr  \left(E^N \right)$. If the state is normalized, then, the spectrum of the matrices $E$ is such that $\{\lambda_{\mu} | ~|\lambda_{\mu}|\le1,  ~1 \le \mu \le D^2 \}$. If the state fulfills the clustering principle\footnote{This principle requires that $\lim_{x \to \infty}{
\la \Psi |\mathcal{O}(x) \mathcal{O}(0) |\Psi \ra}= \la \Psi |\mathcal{O}(x) |\Psi\ra \la \Psi| \mathcal{O}(0) |\Psi \ra$}, then, the maximum eigenvalue $\lambda_1=1$ is unique\cite{Fannes:1990ur}.

In this representation the two point function of any two operators at the site $(i)$ and $(j)$ is written as
\be
\la O^{(i)} O^{(j)} \ra = \tr  \left(E^{N-j+i-1} \tilde{O}^{(i)} E^{j-i-1} \tilde{O}^{(j)} \right)
\ee
where $\tilde{O}=\sum_{\{s\}=1}^d \sum_{\{\tilde{s}\}=1}^d  (A^{s})^* \otimes A^{\tilde{s}} \la s| O | \tilde{s}\ra$. If we consider an infinite chain and taking the diagonal form of $E=|1\ra \la 1|+\sum_{\mu=2}^{D^2}\lambda_{\mu} |\mu\ra \la \mu |$, then,
\be
\begin{split}
\la O^{(i)} O^{(j)} \ra& - \la O^{(i)} \ra \la O^{(j)} \ra  = \la 1| \tilde{O}^{(i)} \left(\sum_{\mu=2}^{D^2} (\lambda_{\mu})^{j-i-1} |\mu\ra \la \mu | \right) \tilde{O}^{(j)} |1\ra \\
&= \sum_{\mu=2}^{D^2} \la 1| \tilde{O}^{(i)} |\mu\ra \la \mu |\tilde{O}^{(j)} |1\ra \left(\frac{\lambda_{\mu}}{|\lambda_{\mu}|}\right)^{j-i-1} e^{-\frac{j-i-1}{\xi_{\mu}}},
\end{split}
\ee
where $\xi_{\mu}=\frac{-1}{\log{|\lambda_{\mu}|}}$ defines the correlation length. From this expression, we see that the long range behavior of the system is implement in the eigenvalues of the transfer matrix $E$, that is invariant under the redefinition of the local basis. An important point that appears about the matrix product state formalism is that any correlation function is expressed as a sum of exponential. This implies that within this formalism to get a non exponential behavior in the correlation, i.e. in the scaling region or at distance between the lattice spacing $a$ and the correlation length $\xi$, $a<x<\xi$, the number of eigenvalues of the matrix $E$ should be as large as the number of coefficients in the Laplace transform of the polynomial decay in the correlation. 

\section{Renormalization group on quantum states.}

In the last section, we have seen that matrix product states is a simple formalism to implement translationally invariant state, that  any expectation value can be calculated in an easy way and, also, they are optimal to minimize the ground state energy. One of the reason for these properties is that the set of local $A^s$ matrices define a transfer matrix between neighbor sites. This fact allows to split the long and short range behavior, i.e. there is an explicit separation of scales. Then, the matrix product state are properly suited to implement the classical Kadanoff\cite{Kadanoff:1967gm} block spin at the quantum level.

\subsection{Scale transformations}

Given a one dimensional quantum state $|\Psi\ra$, it can be decomposed, up to some local unitaries, by a basis $\{|s\ra\}$ of the local Hilbert space and a set of local matrices $\{A^s\}$. In this way, we define an equivalent relation $\sim$ between two states $|\Psi\ra \sim |\tilde{\Psi}\ra$ iff they are related with some local unitary operation $|\Psi\ra = \prod_{\otimes i=1}^N U_i |\tilde{\Psi}\ra$. This equivalent relation is motivated by the fact that long range behavior does not depend on the definition of the local basis. As we have seen, the correlation length is defined by the eigenvalues of the matrix $E=\sum_{s=1}^d (A^s)^*\otimes A^s$ that is independent of the local basis.
 
A scale transformation in the matrix product state is defined by blocking neighbor sites and identifying the local Hilbert space as the space spanned by the states in the block, redefining the $A^s$ matrices and relabeling the position of the blocks:
\be
\tilde{A}^{(s_{2i} s_{2i+1})}_{\alpha\gamma}= \sum_{\beta=1}^D  A^{s_{2i}}_{\alpha \beta} A^{s_{2i+1}}_{ \beta \gamma} = \sum_{s_j=1}^{min(d^2,D^2)} \left(U^{(s_{2i} s_{2i+1})}_{s_j}\right)^{\dagger} \lambda^{s_j} V^{s_j}_{\alpha \gamma}
\ee
where $U$ and $V$ are unitary matrices and $\{\lambda^{s_j}\}$ the coefficients of the singular value decomposition of the block spin $A^{s_{2i}}  A^{s_{2i+1}}$. The first equality in this equation is just the coarse-graining or block spin while the second part takes into account the rescaling of the sites in the chain and the identification of the relevant Hilbert space of the block. It is this second step the origin of the irreversibility in the transformation, because once it is identified the relevant space in the block, any information about the local sites is washed out.

So, the renormalization transformation is implemented by: 
\be
A^s ~\underrightarrow{RG} ~\tilde{A}^s=\lambda^{s} V^{s}
\ee
The dimension of the local Hilbert space in this transformation is always bounded by $D^2$ as it comes from the singular value decomposition of the product of two $A$ matrices. 

As an example of this transformation, we will study how evolve the ground state of the AKLT model for a spin one chain. The AKLT model is defined by an antiferromagnet spin hamiltonian with Heisenberg-like interaction between neighbor sites. It is known that the ground state of this model is given by a valence bond state which is an example of the matrix product state. If we take as a local basis a combination of eigenvectors of the $S^z$ operator, such that,
\be
\begin{split}
|x\ra = \frac{1}{\sqrt{2}} &\left( |+\ra + |-\ra \right);~~|y\ra = \frac{1}{i\sqrt{2}} \left( |+\ra - |-\ra \right);~~|z\ra=|0\ra\\
& S^z | n\ra =n |n \ra, n= \{-1,0,+1\},
\end{split}
\ee
then, the $A^s$ matrices are written in terms of the Pauli matrices,
\be
A^x=\frac{1}{\sqrt{3}} \sigma^x;~~A^y=\frac{1}{\sqrt{3}} \sigma^y;~~A^z=\frac{1}{\sqrt{3}} \sigma^z;
\ee
from which the correlation length of this state is $\xi=\frac{1}{\log{3}}$. In this case, the scale transformation is quite simple due to the properties of the Pauli matrices, $\sigma^{\alpha} \sigma^{\beta} = \delta_{\alpha \beta} \sigma^0 + i\epsilon_{\alpha \beta \gamma} \sigma^{\gamma} $, where $\epsilon_{\alpha \beta \gamma}$ is the Levi-Civita tensor, $\sigma^0$ is the identity matrix and $\{\alpha, \beta, \gamma\}=\{x,y,z\} $. So,
\be
\begin{split}
A^xA^x |xx\ra + A^yA^y|yy\ra + A^zA^z |zz\ra&= \frac{1}{3} \sigma^0 \left(|xx\ra +|yy\ra+|zz\ra \right)\\
A^xA^y |xy\ra + A^yA^x|yx\ra&=\frac{i}{3} \sigma^z \left(|xy\ra -|yx\ra \right)\\
A^yA^z |yz\ra + A^zA^y|zy\ra&=\frac{i}{3} \sigma^x \left(|yz\ra -|zy\ra \right)\\
A^zA^x |xy\ra + A^xA^z|yx\ra&=\frac{i}{3} \sigma^y \left(|zx\ra -|xz\ra \right)
\end{split}
\ee
relabeling the Hilbert space in the block,
\be
\begin{split}
&|\tilde{0}\ra = \frac{1}{\sqrt{3}} \left(|xx\ra +|yy\ra+|zz\ra \right)\\
&|\tilde{x}\ra=\frac{1}{\sqrt{2}} \left(|yz\ra -|zy\ra \right)\\
&|\tilde{y}\ra=\frac{1}{\sqrt{2}} \left(|zx\ra -|xz\ra \right)\\
&|\tilde{z}\ra=\frac{1}{\sqrt{2}} \left(|xy\ra -|yx\ra \right)
\end{split}
\ee
redefining the matrices $A^s$,
\be
\tilde{A}^0= \frac{1}{\sqrt{3}} \sigma^0;~~ \tilde{A}^x=\frac{i\sqrt{2}}{3} \sigma^x;~~\tilde{A}^y=\frac{i\sqrt{2}}{3} \sigma^y;~~\tilde{A}^z=\frac{i\sqrt{2}}{3} \sigma^z.
\ee
In this way, we have an exact representation of the state in a long scale, that, once we forget (project) the local spin state that define the Hilbert space of the block, the irreversibility in the transformation appears. In fact, the new local Hilbert space has physical meaning, they are real quantum state of the system: $|0\ra$ is the singlet state and $|x\ra$, $|y\ra$, $|z\ra$ form the triplet. 

At any point in the scale transformation the local state is written by,
\be
|\Psi(\mu)\ra = \sqrt{1-3 \mu^2} \sigma^0 |0\ra +i \mu \sum_{s=\{x,y,z\}} \sigma^s |s\ra
\ee
with correlation length $\xi=\frac{-1}{\log{|1-4\mu^2|}}$. Then, a step in the RG flow is defined by
\be
|\Psi(\mu)\ra \to |\Psi(\tilde{\mu})\ra, ~~ \left(1-4\mu^2\right)^2= \left(1-4\tilde{\mu}^2\right)
\ee
As a consequence of this analysis, we can seen how the entropy evolves for any given number of sites $L$ and correlation length $\xi=\frac{-1}{\log{|1-4\mu^2|}}$,
\be
\begin{split}
S(L,\mu)=&3 \frac{(1-4\mu^2)^{2^L}-1}{4} \log_2{ \frac{1-(1-4\mu^2)^{2^L}}{4}} \\
&- \frac{1+3(1-4\mu^2)^{2^L}}{4} \log_2{ \frac{1+3(1-4\mu^2)^{2^L}}{4}}
\end{split}
\ee 
whenever $L\to \infty$ or $\xi \to 0$, the entropy goes to $S\to \log_2 4$, see also \cite{Fan:2004ga},this fact is a consequence of the structure of the valence bond state produced by a topological hidden order\cite{Girvin:1989fv} and that it is reflected in a non zero entropy, an infinite entanglement length\cite{Verstraete:2004ps,Verstraete:2004qx,Popp:2005jb} or long range correlation in the string order parameter\cite{Nijs:1989ya}.
 
\subsection{Fixed points in the scale transformation}

Up to now, we have seen how to realize exact renormalization group transformation on quantum states. Usually, if limit cycles\cite{Glazek:2002hq,Morozov:2003ik,LeClair:2004ps} are not considered, this kind of transformations end in a fixed point that characterizes the long range behavior of the state. So, determining the conditions for the fixed point in the transformation and  analyzing  the properties of the quantum state will allow to know better how the low energy properties emerges from a given state.

In the last section, we based the transformation in the $A^s$ matrices but, as we saw, their explicit form depend on the choice of the local basis. Also, we found that the transfer matrix from these matrices $E=\sum_s \left(A^s\right)^* \otimes A^s$ are basis independent. In fact, it is this transfer matrix what is used in the calculation of correlation functions and its eigenvalues determine the long range behavior of the state. The definition of the scale transformation with this operators is even simpler: $\tilde{E}= E^2$, and so, a fixed point in this transformation implies studying the class of operators $\{E_{\infty}\}=\{\lim_{n\to \infty}E^n\}$ or, in other words, analyzing the operators $E$ such that the absolute value of its eigenvalues are zero or one. To carry this program, we will fixed our attention in states that can be decomposed with matrices of dimension $D\le2$ and with the knowledge of completely positive trace-preserving maps\cite{Ruskai:2002sv} we will give a complete classification of fixed points in these transformations.

\begin{enumerate}
\item The first example of fixed point in these RG transformations is given by product states, such that $A^1=1$ and $A^s=0~\forall s\ne 1$. The transformation is given by $\tilde{A}^1=A^1A^1$ and the Hilbert space of the block is one dimensional $|\tilde{1}\ra=|11\ra$. Obviously, $\tilde{E}=E$.

\item GHZ states\cite{Greenberger:1989nl} or cat states is another example of a fixed point. The ground state of the Ising model correspond to this kind of states, if there is no symmetry breaking. So, they can be characterized by $A^1=\frac{\sigma^0 + \sigma^z}{2}$ and $A^2=\frac{\sigma^0 - \sigma^z}{2}$, where $\sigma^0$ is the $2\times 2$ identity matrix and $\sigma^z$ the usual z-Pauli matrix. The scale transformation is performed by $\tilde{A}^1=A^1A^1$ and $\tilde{A}^0=A^0A^0$. The Hilbert space in the block is given by $|\tilde{1}\ra=|11\ra$ and $|\tilde{0}\ra=|00\ra$. It is evident that $\tilde{E}=E$.

\item The third case corresponds to the fixed point in the valence bond phase that turns out to have the same decomposition as in the fixed point of a one dimensional cluster state\cite{Briegel:2000pv}. At the fixed point the matrices appears as: $A^0=\frac{1}{2} \sigma^0, ~A^x=\frac{i}{2} \sigma^x, ~A^y=\frac{i}{2} \sigma^y, ~A^z=\frac{i}{2} \sigma^z$ and the transfer matrix,
\be
E=\frac{1}{2} \begin{pmatrix} 1 & 0 & 0 & 1\\ 0 & 0 & 0 & 0\\0 & 0 & 0 & 0\\1 & 0 & 0 & 1 \end{pmatrix}
\ee 
is such that $\tilde{E}=E^2=E$.

\item The third example completes the cases where $E$ is diagonalizable. For the case where $E$ has a Jordan-block decomposition appear another two cases. When the local Hilbert spaces is two dimensional with matrices,
\be
A^0=\begin{pmatrix} 1& 0 \\ 0 & e^{-i\theta} \end{pmatrix}; ~~~A^1= \begin{pmatrix} 0 & 0 \\ 1 & 0 \end{pmatrix}
\ee
it can be seen that the $A^1$ is nilpotent and $A^0A^1=e^{-i\theta}A^1$. The transfer matrix $E$ is written as,
\be
E=\begin{pmatrix} 1& 0 & 0 & 0 \\ 0 & e^{-i\theta} & 0 & 0 \\ 0 & 0 & e^{i\theta} & 0 \\ 1 & 0 & 0 & 1 \end{pmatrix}.
\ee
Then, the final state will be a combination of $|00...010..00\ra$ vectors where at most one $|1\ra$ appears and in the case $\theta=0$, it is recovered the $W$ state\cite{Dur:2000rp}.

\item The last case, it happens when there are domain walls. Then, the decomposition in terms of the matrices $A^s$ will be given by,
\be
\begin{split}
A^0=&\begin{pmatrix} 0 &0 \\ \cos{\alpha} \sin{\beta} & e^{i\theta}\end{pmatrix}; ~ A^1=\begin{pmatrix} 0& 0\\ \sin{\alpha} & 0\end{pmatrix};\\
&~A^2=\begin{pmatrix} e^{-i\theta} &0\\ \cos{\alpha} \cos{\beta} & 0 \end{pmatrix};
\end{split}
\ee
which have a local Hilbert space of dimension three, but dimension two if $\sin{\alpha}=0$. They fulfill: $A^0A^1= e^{i\theta} A^1$, $A^1A^0=0$, $A^0A^2\propto A^1$, $A^2A^0=0$, $A^1A^2 = e^{-i\theta} A^1$, $A^2A^1=0$, $A^0A^0=e^{i\theta} A^0$, $A^1A^1=0$, $A^2A^2=e^{-i\theta} A^2$, and the transfer matrix $E$ will be
\be
E=\begin{pmatrix} 1 & 0 & 0 & 0 \\ e^{-i\theta} \cos{\alpha} \cos{\beta} & 0 & 0 & 0 \\ e^{i\theta} \cos{\alpha} \cos{\beta} & 0 & 0 & 0 \\ 1 & e^{i\theta} \cos{\alpha} \sin{\beta} & e^{i\theta} \cos{\alpha} \sin{\beta} & 1 \end{pmatrix}
\ee
So, the state is a superposition of terms of the form $|00...00122...22\ra$.
\end{enumerate}

An obvious generalization of the kind of fixed point of a valence bond state and cluster state for $D\times D$ matrices corresponds to a symmetric superposition of $D^2$ orthonormal local states $\{|s\ra, ~ 1\le s \le D^2 \}$, with the generators $T^s$ of $\mathcal{M}_{D\times D}$, such that, $T^s_{\alpha \beta}=1$ iff $D(\alpha-1)+\beta=s$ and zero otherwise. Then, the local state appears as,
\be
|\Psi\ra= \sum_{a=1}^{D^2} \frac{1}{\sqrt{D}} T^s |s\ra.
\ee
with a transfer matrix 
\be
E = \frac{1}{D} \sum_{s=1}^{D^2} \left( T^s \right)^* \otimes T^s,
\ee
which fulfills $\tilde{E}=E^2=E$ and it has just one non null eigenvector. Its entropy is $S=\log_2 D^2$ independent of the number of sites. These kind of states are an example of fixed points in which the long range vacuum has non trivial entanglement properties.

\section{Conclusions and outlook.}

To sum up, this third part uses the matrix product state formalism to describe any quantum state and links, at the quantum level, this formalism with block spin transformation that appears in statistical mechanics. We employ this renormalization group transformation on quantum states to characterize and classify the fixed points of this transformation. Then, the main ideas are
\begin{enumerate} 
\item There exists a description of real space renormalization group transformations within matrix product states.
\item We give a complete classification of fixed points of quantum states in non trivial cases.
\end{enumerate}
As we have seen, this RG protocol has allowed us to obtain states that are fixed point of the scale transformation but have non trivial entanglement properties. Therefore, it is possible, within this framework, to study non trivial low energy sectors of a given theory\cite{Wen:zn}.

\bibliographystyle{alpha}
\bibliography{theref}

\appendix

\chapter{Entanglement and order relations}
\label{entan}

\noindent
\begin{center}
%\resizebox{!}{13.3cm}{\includegraphics{portada/conexiones.eps}}
\end{center}
\hfill

\section{Entanglement. Schmidt decomposition. Order relations}

In this appendix, we will fix our attention to the entanglement properties of pure quantum states in the bipartite scenario\footnote{A pure quantum state is described by a normalized vector, $|\psi\ra$, in a Hilbert space, $\mathcal{H}$. The bipartite scenario appears when the Hilbert space is decomposed in the tensor product of two different Hilbert spaces, i.e. $\mathcal{H}=\mathcal{H}_A \otimes \mathcal{H}_B$.}. This case is particularly well understood in quantum information, see for example\cite{:2001fz,Nielsen:2000ne}. In the next lines, we will follow the work of Nielsen\cite{Nielsen:1998sz}, mainly, and Vidal\cite{Vidal:1999qs,Nielsen:2001nr} or Jonathan and Plenio\cite{Jonathan:1999si}, where they relate entanglement properties of a quantum state to order relations.

In a general scenario, any pure state is entangled if it cannot be written as the tensor product of its parts, i.e. $|\psi\ra \ne |\phi_A\ra \otimes |\phi_B\ra \otimes ... |\phi_N\ra \otimes...$, where $|\psi\ra \in \mathcal{H}= \otimes_i \mathcal{H}_i $, and $|\phi_i\ra \in \mathcal{H}_i$. In the bipartite case, one of the most useful tools to describe entanglement properties is the Schmidt decomposition which is a consequence of the singular value decomposition for any complex matrix.  The Schmidt decomposition says that given any pure state $|\psi\ra \in \mathcal{H}_A \otimes \mathcal{H}_B$, there is always an orthonormal basis $\{|\phi^{(i)}_A\ra\}$ for the space $\mathcal{H}_A$ and another  $\{|\phi^{(i)}_B\ra\}$ for $\mathcal{H}_B$ such that $|\psi\ra = \sum_i \mu_i |\phi^{(i)}_A\ra |\phi^{(i)}_B\ra$ with some non-negative real numbers $\mu_i$ called Schmidt coefficients that fulfill $\sum_i (\mu_i)^2=1$. Then, a necessary and sufficient condition for a state to be entangled is that its Schmidt decomposition has more than one coefficient. 

From this decomposition, a direct consequence is deduced for the reduced density matrix of the subsystems, which is defined by tracing out the other subsystem, i.e. $\rho_A \equiv \tr_B |\psi \ra \la \psi |= \sum_i (\mu_i)^2 |\phi^{(i)}_A\ra \la \phi^{(i)}_A| $ and $\rho_B \equiv \tr_A |\psi \ra \la \psi |= \sum_i (\mu_i)^2 |\phi^{(i)}_B\ra \la \phi^{(i)}_B| $. It can be seen from their definitions that they share the same spectrum of eigenvalues. Then, a pure state will be separable, if the reduced density matrices of its parts are one dimensional projectors (pure states), or in other words, the state of the whole system is entangled if the reduced states of its subsystem are mixed.

The amount of mixing in a density matrix can be measured by several mathematical tools, among them appear the entropy of entanglement and the majorization relations. Both quantities are deeply related and their connections with quantum mechanics and statistical physics are a wide and interdisciplinary area of research. In what follows, we will show their mathematical definitions and sketch their basic properties and different meanings with which they can appear.

The fact that majorization and entropy can compare the degree of mixing between two states imposes an order relation in the  set of density matrices. The meaning of order is a well defined concept in mathematics:  a binary relation $\prec$ in a set $\mathcal{A}$, in our case the set of density matrices, is called a preorder relation in $\mathcal{A}$ if and only if $\forall ~ x,y,z \in \mathcal{A}$: (i) $x \prec x$ and (ii) $x \prec y$ and $y \prec z$ then $x \prec z$. The set $\mathcal{A}$ with the preorder relation, i.e. $(\mathcal{A}, \prec)$ is a preordered set. A preorder relation is a partial order relation if (iii) given $x \prec y$ and $y \prec x$ then $x = y$. So, a partial order relation is a reflexive (condition (i)), transitive (condition (ii)) and anti-symmetric (condition (iii)) relation. In addition to this, a partial ordered set is totally ordered if $\forall ~ x , y \in \mathcal{A}$, it happens that $x\prec y$ or $y \prec x$.

\section{Majorization relations and entropy of entanglement}

Initially, majorization relations were defined in statistics and economy fields to compare two probability distributions\cite{Marshall:1979go, Bhatia:1996yh}. So, given two normalized vectors $x,y \in \mathcal{R}^n$, $y$ majorizes $x$, written $x\prec y$, if the following set of inequalities are fulfilled,
\be
\label{ineq}
\begin{split}
x_1^{\downarrow}\le& y_1^{\downarrow}\\
x_1^{\downarrow}+x_2^{\downarrow}\le & y_1^{\downarrow}+y_2^{\downarrow}\\
\vdots& \\
x_1^{\downarrow}+\cdots +x_{n-1}^{\downarrow}\le &y_1^{\downarrow}+\cdots +y_{n-1}^{\downarrow}\\
x_1^{\downarrow}+\cdots +x_n^{\downarrow} = &y_1^{\downarrow}+\cdots +y_n^{\downarrow},
\end{split}
\ee
where the notation $x_i^{\downarrow}$ corresponds to the components of the vector $x$ written in non-increasing order, i.e. $x_i^{\downarrow} \ge x_i^{\downarrow} ~\forall i<j$ and $1\le i \le n$. 

As an example, it can be seen that for any normalized vector $x\in \mathcal{R}^n$, it always happens that $(\frac{1}{n},\frac{1}{n},\cdots,\frac{1}{n})\prec (x_1^{\downarrow},x_2^{\downarrow},\cdots,x_n^{\downarrow}) \prec (1,0,\cdots,0)$. So, majorization coincides with the intuitive idea that when $x$ is a more chaotic, disorder or mixed probability distribution than $y$, then $x\prec y$. If in a set $\mathcal{A}$, we define an equivalence relation $\sim$ between elements $x,y\in \mathcal{A}$ with the same majorization properties, i.e. $x\sim y$ if $x \prec y$ and $y \prec x$ then majorization defines a partial order in the set $\mathcal{A}$.

The connection between majorization and quantum mechanics can be done using the spectrum of the density matrix as a normalized probability distribution. In fact, given two hermitian matrices $X$ and $Y$, it is defined $X \prec Y$, $X$ is majorized by $Y$, if $\Lambda(X) \prec \Lambda(Y)$, where $\Lambda(X)$ is the spectrum of the eigenvalues of the matrix $X$.

There are other definitions for majorization equivalent to eqs.(\ref{ineq}) that do not depend on the order of the eigenvalues of the density matrix. We will show three results that reveal the close relation between majorization and unitarity. Also, we will see that two density matrices are related by majorization if the more disorder matrix can be obtained from the purer one by a determined set of quantum operations.

Any quantum operation $\mathcal{E}$ acting on a density matrix $\rho$ can be written as $\mathcal{E}(\rho)= \sum_i E_i \rho E_i^{\dagger}$. Where $\{E_i\}$ is a set of operation elements that, for any physical evolution, they must obey the completeness relation, $\sum_i E^{\dagger}_i E_i = \mathbb{I}$ or equivalent any quantum operation is a trace-preserving map between the set of the density matrices, $\tr(\mathcal{E}(\rho))= \tr(\rho)=1$.

A specific set of quantum operation are those ones that has the identity matrix $\mathbb{I}$ as a fixed point, i.e. $\mathcal{E}(\mathbb{I})=\sum_i E_iE^{\dagger}_i=\mathbb{I}$. This kind of operation are called unital operations. The quantum operations that are unital, positive and trace-preserving maps are called doubly stochastic quantum operations. An special case of these ones are the random unitary operations, written $\mathcal{E}(\rho)=\sum_i p_i U_i \rho U_i^{\dagger}$, where $p_i$ is a probability distribution and $U_i$ are unitary matrices.

The following theorem due to Uhlmann characterizes majorization in terms of doubly stochastic and random unitary quantum operations: 

\emph{Uhlmann's theorem}.- Given any pair of hermitian matrices $X$ and $Y$ the following statements are equivalent, (i) $X\prec Y$; (ii) $X$ can be obtained from $Y$ applying a random unitary quantum operation $\mathcal{E}$, i.e. $X= \mathcal{E}(Y)$; (iii) $X$ can be obtained from $Y$ applying a double stochastic quantum operation $\mathcal{E}$, i.e. $X= \mathcal{E}(Y)$.

If we write the equivalent statements for the eigenvalues of the density matrices: (i) $\Lambda(X)\prec \Lambda(Y)$; (ii) $\Lambda(X)$ can be obtained from $\Lambda(Y)$ by a convex combination of permutations of its components, i.e. $\Lambda(X)=\sum_j p_j P_j \Lambda(Y)$, where the real numbers $p_j$ are such that $0 \le p_j \le 1$, $\sum_j p_j =1$ and $\{P_j\}$ is a set of permutation matrices; (iii) $\Lambda(X)$ can be obtained from $\Lambda(Y)$ applying a double stochastic matrix $D$, i.e. $\Lambda(X)= D \Lambda(Y)$.

The fact that two density matrices fulfill majorization relations if they are related by a double stochastic operation characterizes the set of accessible states from a given one. This characterization is a consequence of the next theorem, due to Birkhoff,

\emph{Birkhoff's theorem}.- The set of doubly stochastic matrices is a convex set whose extreme points are permutation matrices.

So, the set $\{x ~|~ x \prec y, ~x,y \in \mathcal{A} \}$ is the convex hull of all points obtained from $y$ by permuting its coordinates.

The third result was shown by Horn and says,

\emph{Horn's lemma}.- Given two vectors $x,y \in \mathcal{R}^n$ with $x \prec y$, there is an orthogonal matrix $u$ such that $x_i= \sum_j |u_{ij}|^2 y_j$. Conversely, given an unitary matrix $u$ such that $x_i= \sum_j |u_{ij}|^2 y_j$ then $x \prec y$.

As we said in the introduction, apart from majorization, there are many other quantities that can measure the degree of mixing in a density matrix and give a notion of order in a probability distribution. Such kind of quantities that preserve the majorization relations are called Schur-convex function. A function $\phi$ defined on $\mathcal{R}^n$ is Schur-convex  if given $x,y \in \mathcal{R}^n$ such that $x \prec y$ then $\phi(x)\le\phi(y)$. 

An example of this set of function is given by any convex function, in particular, the entropy\cite{Ohya:1993tf,Wehrl:1978jn} of a density matrix $\rho$ defined as: $S(\rho)=-\tr{\rho \log_2{\rho}}$. In fact, the entropy is not a Schur-convex function, but concave. So, given two density matrices $\rho$ and $\sigma$ such that $\rho \prec \sigma$, we know by Uhlmann's theorem that $\rho=\sum_i p_i U_i \sigma U_i^{\dagger}$. Then $S(\rho)=S\left(\sum_i p_i U_i \sigma U_i^{\dagger}\right)$. Using the concave property of the entropy gives, $S(\rho)\ge S(\sigma)$.

\section{Entanglement transformation}

Quantum information perspective has brought new insights and meanings to quantities like majorization or entropy. One of the new task arising for the studying of communication and processing of information with quantum mechanic support was the transformation of quantum state using local operations and classical communication (LOCC). In a bipartite system, it is know that a protocol involving LOCC is equivalent to perform a measurement in one of the parties and applied a unitary transformation, that depends on the outcome of the measurement, in the second subsystem. Within this protocol, the question is to characterized the set of states $\{|\phi \ra \}$ that are accessible to a given one $|\psi\ra$, i.e. $|\psi\ra \underrightarrow{LOCC} |\phi \ra$.

In 1998, Nielsen\cite{Nielsen:1998sz} showed that majorization relations determined this set of states for a single copy of the state. His proof goes as follows; from the point of view of the subsystem that realizes the measurement $E_j$, the state will change to $E_j \rho_{\psi} E_j^{\dagger}=p_j \rho_{\phi}$, where $\rho_{\psi}$ is the reduced density matrix before the measurement, $\rho_{\phi}$ is the density matrix after the measurement and $p_j$ is a probability constant. If we used the polar decomposition,
\be
E_j \sqrt{\rho_{\psi}} = \sqrt{E_j \sqrt{\rho_{\psi}} } \sqrt{\sqrt{\rho_{\psi}} E_j^{\dagger} } V_j = \sqrt{p_j \rho_{\phi}} V_j
\ee
with some unitary $V_j$. Then, $\rho_{\psi}=\sum_j p_j V_j^{\dagger} \rho_{\phi} V_j$ and $|\psi\ra \prec |\phi\ra$.
 
In 1996 Bennett, Bernstein, Popescu and Schumacher\cite{Bennett:1995tk}, using the singlet state as the unit of quantum correlation, showed the number $m$ of singlets that can be extracted from an ensemble of $n$ copies of a given bipartite state. The rate of inter-conversion is given by the entropy of entanglement in the asymptotic limit. Their proof is sketched in the next lines. Without loss of generality, let $|\psi\ra=\cos{\theta} |00\ra + \sin{\theta} |11\ra $ be the Schmidt decomposition of a bipartite, two level quantum state, where $\theta \in [0,2\pi)$ is some angle and $\{|0\ra, |1\ra \}$ is an orthonormal local basis for every subsystem. 
We know that the entanglement entropy of this system is given by, $E(|\psi\ra)=-\cos^2{\theta} \log_2 \cos^2{\theta} -\sin^2{\theta} \log_2 \sin^2{\theta}$. If there are $n$ copies of this state then, 
\be
\begin{split}
|\psi\ra^{\otimes n} &= \sum_{x\in[0,1]^n} \left( \cos{\theta} \right)^{n-|x|} \left(\sin{\theta}\right)^{|x|} |x\ra |x\ra \\
&=\sum_{\omega=0}^n \left( \cos{\theta} \right)^{n-\omega} \left(\sin{\theta}\right)^{\omega} \sum_{|x|=\omega} |x\ra |x\ra,
\end{split}
\ee
where $|x|$ corresponds to the number of one's in binary representation. The state
\be
|\omega \ra =\left[ \begin{pmatrix} n \\ \omega \end{pmatrix} \right]^{-1/2} \sum_{|x|=\omega} |x\ra |x\ra
\ee
is a maximally entangled state equivalent to $m=\log_{2}{\begin{pmatrix} n \\ \omega \end{pmatrix}}$ singlets. From the $n$ copies of the initial state, there is a probability $p(\omega)=\begin{pmatrix} n \\ \omega \end{pmatrix} \cos{\theta}^{2(n-\omega)} \sin{\theta}^{2\omega}$ to obtain the maximally entangled state $|\omega \ra$. This probability distribution can be described by a gaussian distribution when $n\to \infty$ with a mean value $n \sin^2{\theta}$ and variance $n \sin^2{\theta} \cos^2{\theta}$. So, on average and in the asymptotic limit,
\be
m=\log_{2}{\begin{pmatrix} n \\ \omega \end{pmatrix}} \simeq \log_{2}{\begin{pmatrix} n \\ n \sin^2{\theta} \end{pmatrix}} \simeq n E(|\psi\ra),
\ee
where it is used the Stirling approximation, $\log_2{n!} \simeq n \log_2 n - n$.

\chapter{Hilbert space in a conformal theory}

\noindent
\begin{center}
%\resizebox{!}{13.3cm}{\includegraphics{portada/agujeronegro.eps}}
\end{center}
\hfill

\section{Introduction and notation}

The behavior of two point correlation function is one of the main properties that characterizes a critical theory. In field theory, the propagator is defined by the vacuum expectation value:
\be
G(t,r) \equiv \la \phi(t,r) \phi(0,0) \ra = \la \phi(r) e^{-i \hat{H} t/\hbar} \phi(0) \ra,
\ee
where in the first equality, the fields are defined in the Heisenberg picture and in the last one in the Sch{\"o}dinger picture. To study its properties, usually, it is performed a Wick rotation in the time direction, in which $it \to \tau$ where $\tau \in \mathbb{R}$. So, the space-time coordinates are mapped into the complex plane where the real part takes into account the euclidean time direction and the imaginary part the coordinate direction. In these lines, we will work in this framework and we will note the space-time surface in complex coordinates as $z=\tau + i r$ and $\bar{z}=\tau - i r$.

In a critical theory, two points correlations transform in a covariant way under a scale transformation $G(r)=b^{-2x} G(r/b)$. From this expression, it is assumed a covariance transformation under dilatations for a kind of operator called scaling operators, $\phi(r)=b^{-x} \phi(r/b)$, where $x$ is the scaling dimension of $\phi$. An extension to other transformations that keep the covariance of scaling operators (or a subset of them), shows that any local combination of translations, rotations and dilatations in the complex plane can still be considered. This set of transformations forms part of a bigger group called conformal. In fact, any conformal transformation can be defined as that coordinate transformation which keeps the metric invariant, up to a local scale factor, i.e., $g'_{\mu, \nu}(r')= \Omega(r) g_{\mu, \nu}(r)$.

Given any infinitesimal local coordinate transformation, $r'^{\mu}=r^{\mu}+\alpha^{\mu}(r)$, the line element defined as $ds^2=g_{\mu \nu} dr^{\mu} dr^{\nu}$ will change to $ds'^2=ds^2+(\partial_{\mu}\alpha_{\nu}+\partial_{\nu}\alpha_{\mu}) dr^{\mu}dr^{\nu}$. If the transformation is conformal, the metric is invariant up to a constant, then $(\partial_{\mu}\alpha_{\nu}+\partial_{\nu}\alpha_{\mu}) = \partial_{\lambda} \alpha^{\lambda} g_{\mu \nu}$, which implies in the flat complex plane, where $g_{\mu \nu}= \delta_{\mu \nu}$ and $ds^2=dzd\bar{z}$, that $\partial_{z}\alpha^{\bar{z}}=\partial_{\bar{z}}\alpha^{z}=0$. Therefore, the conformal group in two dimension is isomorphic to the group of analytic transformation in the complex plane where $z \to \omega(z)$ and $\bar{z}\to \omega(\bar{z})$. For example, a global translation in the plane is represented by $z \to z'=z+a$;  a rotation by an angle $\theta$ is described by $z \to z' = z e^{-i \theta}$; or a global dilatation can be recast as $z \to z' = b z$.

Motivated by the models that appear in spin systems, as an explicit example, we will show how a free fermion theory looks like within this formalism. The Ising model can be characterized by a hamiltonian that reads,
\be
H =  \int d\phi \, \left| \phi \right| \hat{b}^{\dagger}_{\phi} \hat{b}_{\phi} =  \int_{\phi>0} d\phi \,  \phi  \left( \hat{b}^{\dagger}_{\phi} \hat{b}_{\phi} +   \hat{b}^{\dagger}_{-\phi} \hat{b}_{ -\phi} \right).
\ee
with a set of anticommuting operators $\{\hat{b}^{\dagger}_{\phi'}, \hat{b}_{\phi}\}=\delta_{\phi' \phi}$ that evolve in the Heisenberg picture like, $\dot{\hat{b}}_{\phi}=i[H,\hat{b}_{\phi}]=-i|\phi|\hat{b}_{\phi}$ or $\hat{b}_{\phi}(t)=e^{-i|\phi|t}\hat{b}_{\phi}$. It has a vacuum state $|0\ra$ such that $\hat{b}_{\phi} |0\ra = 0 ~\forall \phi$ and the majorana fermions
\be
\begin{split}
&\check{a}_{R,r}=\int_{\phi>0} \frac{d\phi}{\sqrt{2\pi}}  \left( e^{i\phi r} \hat{b}_{\phi}+e^{-i\phi r} \hat{b}^{\dagger}_{\phi} \right),\\
&\check{a}_{L,r}=\int_{\phi>0} \frac{d\phi}{\sqrt{2\pi}} \left( e^{-i\phi r} \hat{b}_{-\phi}+e^{i\phi r} \hat{b}^{\dagger}_{-\phi} \right),\\
&\{\check{a}_{\alpha,r}, \check{a}_{\beta,r'} \} = \delta_{\alpha\beta} ~ \delta(r-r'), \, ~~\, \check{a}_{\alpha,r} \check{a}_{\alpha,r} = \frac{1}{2},
\end{split}
\ee
map the hamiltonian into,
\be
H = \frac{-i }{2}  \int \, dr \, \left( \check{a}_{R,r} \partial_{r} \check{a}_{R,r} -\check{a}_{L,r} \partial_{r} \check{a}_{L,r}  \right).
\ee

The time ordered correlation, i.e. $t_1>t_2$, is written as,
\be
\la \check{a}_{R}(t_1,r_1) \check{a}_{R}(t_2,r_2) \ra = \frac{1}{2\pi} \int_{\phi>0} d\phi ~ e^{i\phi (r_1-t_1)} e^{-i\phi (r_2-t_2)}
\ee
in the complex coordinates where $\tau=it$ with $\tau \in \mathbb{R}$ and $\bar{z}=\tau-ir$, the last equation is recast into
\be
\la \check{a}_{R}(z_1,\bar{z}_1) \check{a}_{R}(z_2,\bar{z}_2) \ra = \frac{1}{2\pi} \int_{\phi>0} d\phi ~ e^{-\phi (\bar{z}_1-\bar{z}_2)}=\frac{1}{2\pi} \frac{1}{\bar{z}_1-\bar{z}_2}
\ee
where $\mathcal{R}e\{\bar{z}_1\}>\mathcal{R}e\{\bar{z}_2\}$. In a similar way, it can be obtained the rest of time ordered correlators,
\be
\begin{split}
&\la \check{a}_{L}(z_1,\bar{z}_1) \check{a}_{L}(z_2,\bar{z}_2) \ra =\frac{1}{2\pi} \frac{1}{z_1-z_2}\\
&\la \check{a}_{R}(z_1,\bar{z}_1) \check{a}_{L}(z_2,\bar{z}_2) \ra =\la \check{a}_{L}(z_1,\bar{z}_1) \check{a}_{R}(z_2,\bar{z}_2) \ra =0,
\end{split}
\ee
which yields the scaling dimensions $x=\frac{1}{2}$ for the operators $\check{a}_{R}$ and $\check{a}_{L}$. Also, it can be obtained the correlator with the fields $\partial_{z}\check{a}_{L}(z) $ or $\partial_{\bar{z}}\check{a}_{R}(\bar{z})$,
\be
\begin{split}
\la \partial_{\bar{z}_1} \check{a}_{R}(z_1,\bar{z}_1) \check{a}_{R}(z_2,\bar{z}_2) \ra &= \frac{-1}{2\pi} \frac{1}{\left(\bar{z}_1-\bar{z}_2 \right)^2}\\
\la \partial_{z_1} \check{a}_{L}(z_1,\bar{z}_1) \check{a}_{L}(z_2,\bar{z}_2) \ra &=\frac{-1}{2\pi} \frac{1}{\left(z_1-z_2\right)^2} \\
\la \partial_{\bar{z}_1} \check{a}_{R}(z_1,\bar{z}_1) \partial_{\bar{z}_2}  \check{a}_{R}(z_2,\bar{z}_2) \ra &= \frac{-1}{\pi} \frac{1}{\left(\bar{z}_1-\bar{z}_2 \right)^3}\\
\la \partial_{z_1} \check{a}_{L}(z_1,\bar{z}_1) \partial_{z_2} \check{a}_{L}(z_2,\bar{z}_2) \ra& =\frac{-1}{\pi} \frac{1}{\left(z_1-z_2\right)^3}
\end{split}
\ee
 
\section{Conformal symmetry}

Following Belavin, Polyakov and Zamalodchikov\cite{Belavin:1984vg}, a conformal invariant theory can be defined in an axiomatical way:

\begin{enumerate}
\item The theory is described by the correlation functions of a set (in general infinite) of local scaling operators $\{\phi(z,\bar{z})\}$.
\item This set of operators $\{\phi(z,\bar{z})\}$ is supposed to be complete in the sense that they close the operator algebra $\phi_i(z) \phi_j(0) = \sum_k C_{ij}^k(z) \phi_k(0)$ with structure constants $C_{ij}^k (z)$.
\item There is a subset of operators called primary which transform under any conformal transformation $\omega(z)$, global or local, covariantly,
\be
A_{\Delta, \bar{\Delta}}(z,\bar{z}) \to \left(\frac{\partial \omega}{\partial z}\right)^{\Delta} \left(\frac{\partial \bar{\omega}}{\partial \bar{z}}\right)^{\bar{\Delta}} A_{\Delta, \bar{\Delta}}(\omega,\bar{\omega}) 
\ee
where the real numbers $(\Delta,\bar{\Delta})$ are called complex scaling dimensions or conformal weights.
\item Any local operator can be written as a linear combination of the primary operators and their derivatives.
\item The vacuum is invariant under global conformal transformations.
\item The generators of conformal transformations are given by the components of the energy-momentum tensor $T_{\mu \nu}(z)$, characterized by the Virasoro algebra.
\end{enumerate}

Before showing in more detail these postulates, it has to be fixed the time direction in the complex plane. This fact is done using concentric circles around the origin of the complex plane as the equal time directions, this choice is called radial ordering. For instance, we can parametrize a finite model of $N$ sites by the coordinates $z=\tau+ir$, identifying the extreme points in the space direction, i.e. $r+N=r$, so, the model is placed in a cylindrical geometry. Also, we can place the system in the entire complex plane using the conformal transformation, $\omega=e^{\frac{2\pi z}{N}}= e^{\frac{2\pi}{N} (\tau + i r)}$ which maps the infinite past $\tau \to -\infty$ to the origin of the plane and the equal time directions are organized in concentric lines around this point. With this mapping, the time reversal operation $\tau \to -\tau$ is performed by $\omega \to \frac{1}{\omega^*}$, and time translations in the cylinder $\tau \to \tau + a$ corresponds to dilatations in the plane $\omega \to e^a \omega$.

With the postulates and the radial quantization, we will see how global conformal invariance fixes the functional form of the propagator between any two primary operator. Under any conformal transformation $z \to \omega(z)$, the two point function with $|z_1|>|z_2|$ will change to
\be
\begin{split}
\la A_{\Delta_1, \bar{\Delta}_1}(z_1,\bar{z}_1) A_{\Delta_2, \bar{\Delta}_2}(z_2,\bar{z}_2) \ra& = \left(\frac{\partial \omega}{\partial z_1}\right)^{\Delta_1} \left(\frac{\partial \bar{\omega}}{\partial \bar{z}_1}\right)^{\bar{\Delta}_1}  \left(\frac{\partial \omega}{\partial z_2}\right)^{\Delta_2} \cdot \\
\cdot \left(\frac{\partial \bar{\omega}}{\partial \bar{z}_2}\right)^{\bar{\Delta}_2} & \la A_{\Delta_1, \bar{\Delta}_1}(\omega_1,\bar{\omega}_1) A_{\Delta_2,\bar{\Delta}_2}(\omega_2,\bar{\omega}_2) \ra,
\end{split}
\ee
in the case of an infinitesimal conformal transformation, $z \to z+\alpha(z)$, primary operators transform as 
\be
\begin{split}
&\delta_{\alpha} A_{\Delta, \bar{\Delta}}(z,\bar{z})=\\
& =\left(1+\partial_z \alpha \right)^{\Delta} \left(1+\partial_{\bar{z}} \bar{\alpha} \right)^{\bar{\Delta}} A_{\Delta, \bar{\Delta}}\left(z+\alpha(z),\bar{z}+\bar{\alpha}(z)\right) -A_{\Delta, \bar{\Delta}}(z,\bar{z})  \\
&=\left[ \left( \Delta \partial_{z}\alpha + \alpha \partial_{z} \right)+\left(\bar{\Delta} \partial_{\bar{z}}\bar{\alpha} + \bar{\alpha} \partial_{\bar{z}} \right) \right] A_{\Delta, \bar{\Delta}}(z,\bar{z}) + O(\alpha^2),
\end{split}
\ee
global transformations do not change the two point function, then
\be
\begin{split}
&\delta_{\alpha} \la  A_{\Delta_1, \bar{\Delta}_1}(z_1,\bar{z}_1) A_{\Delta_2, \bar{\Delta}_2}(z_2,\bar{z}_2) \ra = \\
&=\la \delta_{\alpha} A_{\Delta_1, \bar{\Delta}_1}(z_1,\bar{z}_1) A_{\Delta_2, \bar{\Delta}_2}(z_2)\ra + \la  A_{\Delta_1, \bar{\Delta}_1}(z_1) \delta_{\alpha} A_{\Delta_2, \bar{\Delta}_2}(z_2,\bar{z}_2)\ra \\
&= \left(\sum_{n=1,2} \Delta_n \partial_{z_n}\alpha + \alpha \partial_{z_n} +\bar{\Delta}_n \partial_{\bar{z}_n}\bar{\alpha} + \bar{\alpha} \partial_{\bar{z}_n} \right) \cdot \\ 
&~~~~~~~~~~~~~~~~~~~~~~~~~~~~~~~~~~~~~~~~\cdot \la A_{\Delta_1, \bar{\Delta}_1}(z_1,\bar{z}_1)  A_{\Delta_2, \bar{\Delta}_2}(z_2,\bar{z}_2) \ra =0
\end{split}
\ee
Translational invariance, i.e. $\alpha=1$ and $\bar{\alpha}=1$, shows that the coordinate dependance should be like $z_{12}=z_1-z_2$ and $\bar{z}_{12}=\bar{z}_1-\bar{z}_2$. Scale invariance, i.e. $\alpha=z$ and $\bar{\alpha}=\bar{z}$, fixes the functional dependance,
\be
\la A_{\Delta_1, \bar{\Delta}_1}(z_1,\bar{z}_1) A_{\Delta_2, \bar{\Delta}_2}(z_2,\bar{z}_2) \ra = \frac{C_{12}}{z_{12}^{\Delta_1 +\Delta_2} \bar{z}_{12}^{\bar{\Delta}_1 +\bar{\Delta}_2} },
\ee
where $C_{12}$ is a constant that can be determined by normalization. Finally, using $\alpha=z^2$ and $\bar{\alpha}=\bar{z}^2$ implies,
\be
\la A_{\Delta_1, \bar{\Delta}_1}(z_1,\bar{z}_1) A_{\Delta_2, \bar{\Delta}_2}(z_2,\bar{z}_2) \ra =
\begin{cases}
z_{12}^{-2\Delta} \bar{z}_{12}^{-2\bar{\Delta}} & \Delta_1=\Delta_2=\Delta \\
0 & \Delta_1\neq \Delta_2
\end{cases}
\ee
From this expression, the scaling dimensions can be defined in terms of the conformal weights $x=\Delta+\bar{\Delta}$ and the spin of the operator\footnote{Although, it is called spin, it has no relation with the spin property in quantum mechanic system.}, $s=\Delta-\bar{\Delta}$, which means that the two points correlator behaves as $G(z,\bar{z})=|z|^{-2x} \left( \frac{\bar{z}}{z} \right)^{s}$.

A conformal theory postulates that the operators  $A(z_1)$ and $A(z_2)$ are related via the operator algebra, in the limit $z_1 \to z_2$, the product of these operators can be characterized with the most divergent parts, so, it is usually defined the product expansion at the level of operators: 
\be
A_{\Delta_1, \bar{\Delta}_1}(z_1,\bar{z}_1) A_{\Delta_2, \bar{\Delta}_2}(z_2,\bar{z}_2) \sim \frac{C_{12}}{z_{12}^{\Delta_1 +\Delta_2} \bar{z}_{12}^{\bar{\Delta}_1 +\bar{\Delta}_2} }
\ee

As we saw, in the case of a free fermion $\Delta_L=\bar{\Delta}_R=\frac{1}{2}$ and  $\Delta_R=\bar{\Delta}_L=0$ and the operator product expansion (OPE) appears as,
\be
\begin{split}
&\check{a}_{R}(z_1,\bar{z}_1) \check{a}_{R}(z_2,\bar{z}_2)  \sim \frac{1}{2\pi} \frac{1}{\bar{z}_1-\bar{z}_2} \\
& \check{a}_{L}(z_1,\bar{z}_1) \check{a}_{L}(z_2,\bar{z}_2)  \sim \frac{1}{2\pi} \frac{1}{z_1-z_2}
\end{split}
\ee

Another important OPE in conformal theories is the expansion of the operators with the energy-momentum tensor. In the case of the fermionic model, the components of this tensor, in normal ordering, are written as,
\be
\begin{split}
T(z)&=- \pi: \check{a}_{L}(z) \partial_z  \check{a}_{L}(z): \\
&= - \pi \lim_{z\to \omega} \left( \check{a}_{L}(z) \partial_{\omega}  \check{a}_{L}(\omega) - \la   \check{a}_{L}(z) \partial_{\omega}  \check{a}_{L}(\omega) \ra \right) \\ 
\bar{T}(\bar{z})& =-\pi : \check{a}_{R}(\bar{z}) \partial_{\bar{z}}  \check{a}_{R}(\bar{z}):\\
&=-\pi  \lim_{\bar{z} \to \bar{\omega}}  \left(  \check{a}_{R}(\bar{z}) \partial_{\bar{\omega}}  \check{a}_{R}(\bar{\omega}) - \la \check{a}_{R}(\bar{z}) \partial_{\bar{\omega}}  \check{a}_{R}(\bar{\omega}) \ra \right) 
\end{split}
\ee
Because this theory is gaussian, the Wick theorem can be applied in the product of these operators. Taking into account the contractions between pairs, with the sign that comes from the permutation of the fields,
\be
\begin{split}
&T(z)\check{a}_{L}(\omega)=-\pi: \check{a}_{L}(z) \partial_z  \check{a}_{L}(z): \check{a}_{L}(\omega) \sim \frac{1}{2} \frac{\check{a}_{L}(\omega)}{\left(z-\omega \right)^2}+ \frac{\partial_{\omega}  \check{a}_{L}(\omega)}{z-\omega} \\
&\bar{T}(\bar{z}) \check{a}_{R}(\bar{\omega}) = -\pi : \check{a}_{R}(\bar{z}) \partial_{\bar{z}}  \check{a}_{R}(\bar{z}): \check{a}_{R}(\bar{\omega}) \sim \frac{1}{2} \frac{\check{a}_{R}(\bar{\omega})}{\left(\bar{z}-\bar{\omega} \right)^2}+ \frac{\partial_{\bar{\omega}}  \check{a}_{R}(\bar{\omega})}{\bar{z}-\bar{\omega}}, 
\end{split}
\ee
and the expansion of the energy momentum tensor with itself is written 
\be
\begin{split}
T(z)T(\omega)&=\pi^2 : \check{a}_{L}(z) \partial_z  \check{a}_{L}(z): : \check{a}_{L}(\omega) \partial_{\omega}  \check{a}_{L}(\omega): \\
&\sim \frac{1}{4} \frac{1}{(z-\omega)^4} + \frac{2T(\omega)}{(z-\omega)^2} + \frac{\partial_{\omega} T(\omega)}{(z-\omega)}.
\end{split}
\ee
in a similar way it can be done with the right mover or antiholomorphic terms.

\section{Virasoro algebra}

With the radial ordering in the complex plane, it is straightforward to get equal time commutators between two operators. For instance, given two local fields $\phi_1(z)$ and $\phi_2(\omega)$, the radial ordering corresponds to $\phi_1(z) \phi_2(\omega)$ if $|z| > |\omega|$ and $\phi_2(\omega) \phi_1(z)$ if $|\omega| > z$. Then, the contour integral
\be
\oint_{\omega} dz \, ~ \phi_1(z) \phi_2(\omega) 
\ee
is equivalent to the commutator $[\oint dz \phi_1(z),  \phi_2(\omega) ]$, i.e.
\be
\begin{split}
\oint_{\omega} dz \, ~ \phi_1(z) \phi_2(\omega)& = \oint_{|z| > |\omega|} dz \, \phi_1(z) \phi_2(\omega) -  \oint_{|z| < |\omega|} dz \,   \phi_2(\omega) \phi_1(z) \\
&= [\oint dz \phi_1(z),  \phi_2(\omega) ]
\end{split}
\ee
\begin{figure}[!ht]
\begin{center}
%\resizebox{!}{3.0cm}{\includegraphics{grafic/contor.eps}}
\caption[Equal time commutator in the complex plane and contour integrals]{Equivalence between equal time commutator in the complex plane and contour integrals.}
\end{center}
\end{figure}

In the next lines, we will see that knowing the operator product expansion of the energy momentum tensor is equivalent to knowing the commutator of the operator with the generator of the conformal transformations.

In the last section, we saw how primary operators change under a conformal transformation. A way to analyze any transformation is studying its generators, so if an operator changes under an infinitesimal transformation, this change is expressed as a commutator of the operator with the generator of the transformation, i.e., $\delta_{\alpha} A_{\Delta, \bar{\Delta}}(z,\bar{z}) =  \left[ Q_{\alpha} , A_{\Delta, \bar{\Delta}}(z,\bar{z}) \right]$. For a general Lie group with generators $g_i$, an infinitesimal transformation is given by $\epsilon g_i$; in a conformal transformation $r'^{\mu}=r^{\mu}+\alpha^{\mu}(r)$, because the small parameter is a function of the coordinates, a general infinitesimal transformation is postulated as a sum in the complex plane, $Q_{\alpha}=  \frac{1}{2\pi}\int T_{\mu\nu}(r) \partial^{\mu}\alpha^{\nu}(r) d^{2}r$, in analogy with the elasticity theory, which defines the energy-momentum tensor or stress tensor. In fact, $Q_{\alpha}$ is the change in the action when an infinitesimal conformal transformation is applied and obviously this expansion is valid only if the theory is supposed local. 

From the definition and the invariance of the vacuum under global conformal transformations, it implies that:
\begin{enumerate}
\item Translational invariance, i.e. $\alpha^{\nu}(r)$ constant, is trivially obtained
\item Rotational invariance, i.e. $\partial^{\mu}\alpha^{\nu}(r) = -\partial^{\nu}\alpha^{\mu}(r)$, implies that $T_{\mu\nu}(r) = T_{\nu\mu} (r)$, so the energy-momentum tensor is symmetric
\item Scale invariance, i.e. $\partial^{\mu}\alpha^{\nu}(r) =  \delta_{\mu \nu}$, made the stress tensor traceless, $\sum_{\mu} T_{\mu\mu} = 0$
\end{enumerate}

Using the fact that the energy-momentum tensor is symmetric, traceless and a conserved quantity, i.e. $\partial^{\mu} T_{\mu \nu}=0$ , it can be defined two independent component,
\be
T_{zz} = T_{11} -T_{22}- 2i T_{12}, ~~~~~~ \bar{T}_{\bar{z} \bar{z}} = T_{11} -T_{22} + 2i T_{12},
\ee
where $\partial_{\bar{z}} T_{zz}=\partial_{z} \bar{T}_{\bar{z} \bar{z}}=0$ due to the energy-momentum conservations, then $T_{zz}=T(z)$ and $\bar{T}_{\bar{z} \bar{z}} = \bar{T}(\bar{z} )$. Applying the Gauss theorem, the variation in the action can be recast into,
\be
\begin{split}
Q_{\alpha}&=  \frac{1}{2\pi} \int \partial^{\mu} \left( T_{\mu\nu}(r) \alpha^{\nu}(r)\right) d^{2}r \\
&= \frac{1}{2 \pi i} \oint dz \, \alpha(z) T(z) + \frac{1}{2 \pi i} \oint d\bar{z} \, \alpha (\bar{z}) \bar{T}(\bar{z}) 
\end{split}
\ee

Finally, the variation of any primary operator can be obtained as
\be
\begin{split}
&\delta_{\alpha} A_{\Delta, \bar{\Delta}}(\omega,\bar{\omega}) =\\
&= \frac{1}{2\pi i} \oint   \left[  dz \alpha(z) T(z),A_{\Delta, \bar{\Delta}}(\omega,\bar{\omega}) \right] + \left[d\bar{z} \alpha(\bar{z}) \bar{T}(\bar{z}), A_{\Delta, \bar{\Delta}}(\omega,\bar{\omega}) \right]\\
&= \frac{1}{2\pi i} \oint_{\omega}  dz ~\alpha(z) T(z)A_{\Delta, \bar{\Delta}}(\omega,\bar{\omega})  + \frac{1}{2\pi i} \oint_{\bar{\omega}}  d\bar{z}~ \alpha(\bar{z}) \bar{T}(\bar{z}) A_{\Delta, \bar{\Delta}}(\omega,\bar{\omega}) \\
&=\left( \left( \Delta \partial_{\omega}\alpha + \alpha \partial_{\omega} \right)+\left(\bar{\Delta} \partial_{\bar{\omega}}\bar{\alpha} + \bar{\alpha} \partial_{\bar{\omega}} \right) \right) A_{\Delta, \bar{\Delta}}(\omega,\bar{\omega})
\end{split}
\ee
So, applying the residue theorem,
\be
\begin{split}
&T(z) A_{\Delta}(\omega) \sim \frac{\Delta}{(z-\omega)^2} A_{\Delta}(\omega) + \frac{1}{z-\omega} \partial_{\omega} A_{\Delta}(\omega)\\
&\bar{T}(\bar{z}) A_{ \bar{\Delta}}(\omega,\bar{\omega}) \sim \frac{\bar{\Delta}}{(\bar{z}-\bar{\omega})^2}A_{\bar{\Delta}}(\bar{\omega}) + \frac{1}{(\bar{z}-\bar{\omega})} \partial_{\bar{\omega}} A_{\bar{\Delta}}(\omega,\bar{\omega})  
\end{split}
\ee
or writing the commutation relations,
\be
\frac{1}{2\pi i} \left[ T(z),A_{\Delta}(\omega) \right]= \delta(z-\omega) \partial_{\omega} A_{\Delta}(\omega) -\Delta \partial_z \delta(z-\omega)A_{\Delta}(\omega)
\ee
and in a similar way for the antiholomorphic part. These equations match with the OPE we obtained for the free fermion model if $\Delta_L=\bar{\Delta}_R=\frac{1}{2}$ and $\Delta_R=\bar{\Delta}_L=0$.

%Several direct consequence appears after the definition of the energy-momentum tensor. Integrating by parts the change in the action, a term proportional to $\int \partial^{\mu}T_{\mu\nu}  \alpha^{\nu} d^{2}r$ appears, if the result does not depend in the details of the arbitrary function $\alpha^{\nu} $, then $\partial^{\mu}T_{\mu\nu} =0$ and the energy-momentum tensor is a conserved quantity.

%Using these equation to get the change of the two point function after an infinitesimal conformal transformation $z \to z+\alpha(z)=z (1+ \epsilon)$ gives,
%\be
%\la A_{\Delta, \bar{\Delta}}(z_1,\bar{z}_1) A_{\Delta, \bar{\Delta}}(z_2,\bar{z}_2)\ra_{S}= (1+ \epsilon)^{2\Delta} (1+\bar{\epsilon})^{2\bar{\Delta}} \la A_{\Delta, \bar{\Delta}}(z_1,\bar{z}_1) A_{\Delta, \bar{\Delta}}(z_2,\bar{z}_2)\ra_{S+\delta S},
%\ee 
%expanding the r.h.s. in perturbation to order $\epsilon$ this relation is recast into,
%\be
%\begin{split}
%&\la \delta A_{\Delta, \bar{\Delta}}(z_1,\bar{z}_1) A_{\Delta, \bar{\Delta}}(z_2,\bar{z}_2)\ra + \la  A_{\Delta, \bar{\Delta}}(z_1,\bar{z}_1) \delta A_{\Delta, \bar{\Delta}}(z_2,\bar{z}_2)\ra =\\
%&=- \frac{1}{2\pi}\int d^{2}r  \, \partial^{\mu}\alpha^{\nu}(r) \la T_{\mu\nu}(r) A_{\Delta, \bar{\Delta}}(z_1,\bar{z}_1) A_{\Delta, \bar{\Delta}}(z_2,\bar{z}_2)\ra
%\end{split}
%\ee

A remarkable fact about the OPE of the energy-momentum tensor is that it does not appear as the one for primary operators, as we saw in the model of the free fermion. In a general case, the OPE of $T(z)$ reads,
\be
\begin{split}
&T(z)T(\omega)\sim\frac{1}{2}\frac{c}{(z-\omega)^4}+\frac{2T(\omega)}{(z-\omega)^2} + \frac{\partial_{\omega} T(\omega)}{(z-\omega)} \\
&\bar{T}(\bar{z})\bar{T}(\bar{\omega})\sim\frac{1}{2} \frac{\bar{c}}{(\bar{z}-\bar{\omega})^4}+\frac{2\bar{T}(\bar{\omega})}{(\bar{z}-\bar{\omega})^2} + 
\frac{\partial_{\bar{\omega}} \bar{T}(\bar{\omega})}{(\bar{z}-\bar{\omega})}
\end{split}
\ee 
where the coefficient of the first term in the expansion is a c-number that fixes the universality of the model, it is called the central charge or conformal anomaly. In the fermion example, $c=\bar{c}=\frac{1}{2}$. The second term in the OPE tells us that the energy-momentum tensor $T(z)$ has dimensions $(\Delta,\bar{\Delta})=(2,0)$ and $T(\bar{z})$ has dimensions $(\Delta,\bar{\Delta})=(0,2)$. Then, the two point expectation value of the stress tensor is written,
\be
\begin{split}
&\la T(z)T(\omega) \ra = \frac{1}{2}\frac{c}{(z-\omega)^4} \\
&\la \bar{T}(\bar{z})\bar{T}(\bar{\omega}) \ra = \frac{1}{2} \frac{\bar{c}}{(\bar{z}-\bar{\omega})^4},
\end{split}
\ee
and the variation of the stress tensor under infinitesimal conformal transformations reads,
\be
\begin{split}
\delta_{\alpha} T(\omega)&= \frac{1}{2\pi i} \oint_{\omega}  dz ~\alpha(z) T(z)T(\omega) \\
&= \frac{c}{12} \partial^3_{\omega} \alpha(\omega) + 2 T(\omega) \partial_{\omega} \alpha(\omega) + \alpha(\omega) \partial_{\omega} T(\omega)
\end{split}
\ee
or writing it as the commutator of the energy density,
\be
\frac{1}{2\pi i}\left[ T(z), T(\omega) \right]=\delta(z-\omega)\partial_{\omega}T(\omega)-2\partial_z\delta(z-\omega) T(\omega)+\frac{c}{6}\partial_z^3 \delta(z-\omega)
\ee
for finite transformations, it is recast into,
\be
\tilde{T}(\omega)= \left( \frac{dz}{d\omega} \right)^2 T(z) + \frac{c}{12} \{z;\omega\}
\ee
where $\{z;\omega\}$ is the Schwarzian derivative,
\be
\{z;\omega\}=\frac{d^3z/d\omega^3}{dz/d\omega} - \frac{3}{2} \left( \frac{d^2z/d\omega^2}{dz/d\omega} \right)^2
\ee

Once we know how the energy-momemtum tensor acts on primary fields and their OPE, it is useful to expand the tensor and fields in Laurent series, such that
\be
\begin{split}
&A_{\Delta,\bar{\Delta}}(z,\bar{z})=\sum_{m\in\mathcal{Z}}\sum_{n\in\mathcal{Z}} z^{-m-\Delta} \bar{z}^{-n-\bar{\Delta}} A_{m,n} \\
&A_{m,n}=\oint \frac{dz}{2\pi i}  \oint  \frac{d\bar{z}}{2\pi i}  z^{m+\Delta-1} \bar{z}^{n+\bar{\Delta}-1} A_{\Delta,\bar{\Delta}}(z,\bar{z}) \\
&T(z)=\sum_{n\in\mathcal{Z}}  z^{-m-2} L_n; ~~~ L_n= \oint \frac{dz}{2\pi i} z^{n+1} T(z) \\
&\bar{T}(\bar{z})=\sum_{n\in\mathcal{Z}} \bar{z}^{-m-2} \bar{L}_n; ~~~ \bar{L}_n= \oint \frac{d\bar{z}}{2\pi i} \bar{z}^{n+1} \bar{T}(\bar{z}),
\end{split}
\ee
where the quantum generators $L_n$ obey the Virasoro algebra,
\be
\begin{split}
&\left[ L_n, L_m \right] =(n-m) L_{n+m} + \frac{c}{12} n (n^2 -1 )\delta_{n+m,0};\\
&\left[ \bar{L}_n, \bar{L}_m \right] =(n-m) \bar{L}_{n+m} + \frac{\bar{c}}{12} n (n^2 -1 )\delta_{n+m,0};\\
&\left[ L_n, \bar{L}_m \right]=0.
\end{split}
\ee
The primary fields and the $L_n$ generators are related through the following commutators
\be
\begin{split}
[ L_n, &A_{\Delta,\bar{\Delta}}(\omega,\bar{\omega}) ]  = \\
&=\Delta (n+1) \omega^n A_{\Delta,\bar{\Delta}}(\omega,\bar{\omega}) + \omega^{n+1} \partial_{\omega}A_{\Delta,\bar{\Delta}}(\omega,\bar{\omega})~~~(n\ge-1)\\
[ \bar{L}_n, &A_{\Delta,\bar{\Delta}}(z,\bar{z}) ] =\\
&= \bar{\Delta} (n+1) \bar{\omega}^n A_{\Delta,\bar{\Delta}}(\omega,\bar{\omega}) + \bar{\omega}^{n+1} \partial_{\bar{\omega}}A_{\Delta,\bar{\Delta}}(\omega,\bar{\omega})~~~(n\ge-1)
\end{split}
\ee

A direct application of these equations, and specially important in quantum spin models, is the relation between the energy density defined in the complex plane and the one that appears in a chain of N sites. We know that the transformation that maps the plane into the cylinder is given by: $\omega= \tau + ir =\frac{N}{2\pi}\log{z}$, then
\be
\begin{split}
T_{cyl}(\omega)&= \left(\frac{2\pi}{N}\right)^2 \left[ z^2 T_{pla}(z) - \frac{c}{24} \right] \\
& =  \left(\frac{2\pi}{N}\right)^2 \sum_{n\in\mathcal{Z}} \left(L_n -\frac{c}{24} \delta_{n,0} \right) e^{-2\pi n\omega / N}
\end{split}
\ee
which implies that $\la T_{cyl}(\omega) \ra=  - \frac{c}{24}\left(\frac{2\pi}{N}\right)^2$, so the central charge can be measured from finite size correction in a quantum chain. Another important point is that the generator of the dilatations, i.e. $\alpha(z)=z$, in the plane is the zero component in the energy momentum tensor, $L_0 = \oint \frac{dz}{2\pi i} z T_{pla}(z)$, and we know, within the radial ordering, that the dilatations in the plane are related with time translation in the cylinder. Then, the quantum hamiltonian of a conformal invariant system in one dimension with $N$ sites is written as
\be
H= \left(\frac{2\pi}{N}\right)  \left(L_0+ \bar{L}_0 -\frac{c+\bar{c}}{24} \right).
\ee
So, diagonalize the quantum hamiltonian is equivalent to find the eigenvalues and eigenvectors of $L_0+ \bar{L}_0$.

\section{Hilbert space in conformal theories}

In the last section, we have seen that the eigenstates of $L_0$ corresponds to the eigenstates of the quantum hamiltonian. The vacuum in a theory is defined as the state with minimum energy, then, in theories where the energy is bounded from below, it can be fixed to zero and
\be
L_0|0\ra = \bar{L}_0|0\ra=0.
\ee
States created by the action of a primary field into the vacuum in the infinite past, are called highest weight,
\be
|\Delta, \bar{\Delta} \ra = \lim_{\tau \to - \infty} A_{\Delta,\bar{\Delta}} (\tau, r) |0\ra = \lim_{z,\bar{z} \to 0} A_{\Delta,\bar{\Delta}} (z,\bar{z}) |0\ra,
\ee
the adjoint of this state is defined by,
\be
\begin{split}
&\la \Delta, \bar{\Delta}| = |\Delta, \bar{\Delta} \ra^{\dagger}=\left(  \lim_{z,\bar{z} \to 0} A_{\Delta,\bar{\Delta}} (z,\bar{z}) |0\ra \right)^{\dagger} \\
&=  \lim_{z,\bar{z} \to 0}  \la 0 | \left(A_{\Delta,\bar{\Delta}} (z,\bar{z}) \right)^{\dagger} \equiv \lim_{z,\bar{z} \to 0}  \la 0 | A_{\Delta,\bar{\Delta}} (\frac{1}{z},\frac{1}{\bar{z}}) \frac{1}{z^{2\Delta}} \frac{1}{\bar{z}^{2\bar{\Delta}}}
\end{split}
\ee
which gives
\be
\begin{split}
\la \Delta, \bar{\Delta}|\Delta, \bar{\Delta} \ra& =  \lim_{z,\bar{z} \to 0} \frac{1}{z^{2\Delta}} \frac{1}{\bar{z}^{2\bar{\Delta}}}  \la 0 | A_{\Delta,\bar{\Delta}} (\frac{1}{z},\frac{1}{\bar{z}}) A_{\Delta,\bar{\Delta}} (0,0) |0\ra \\
&=\lim_{\omega,\bar{\omega} \to \infty} \omega^{2\Delta} \bar{\omega}^{2\bar{\Delta}}  \la 0 | A_{\Delta,\bar{\Delta}} (\omega,\bar{\omega}) A_{\Delta,\bar{\Delta}} (0,0) |0\ra=1.
\end{split}
\ee
In the case of the energy tensor,
\be
T^{\dagger}(z)=\sum_n \frac{L_n^{\dagger}}{\bar{z}^{n+2}};~~~~~T(1/\bar{z})\frac{1}{\bar{z}^4}=\sum_n \frac{L_n}{\bar{z}^{-m+2}},
\ee
then $L_m^{\dagger}=L_{-m}$. From the OPE
\be
T(z) A_{\Delta} (0) = \sum_{n=-\infty}^{\infty} z^{-n-2} L_nA_{\Delta,\bar{\Delta}} (0,0)  \sim \frac{\Delta}{z^2} A_{\Delta}(0) + \frac{1}{z} \partial A_{\Delta}(0)
\ee
we know that,
\be
L_0|\Delta \ra = \Delta |\Delta \ra; ~~~~~~~ L_n|\Delta\ra = 0 ~\forall n>0.
\ee
We also saw that
\be
\left[L_0, L_n \right] = -n L_n
\ee
then, if $|\Delta \ra$ is an eigenvector of $L_0$ with eigenvalue $\Delta$, $L_n|\Delta \ra$ is an eigenstate with eigenvalue $(\Delta-n)$. So, $L_n$ acts as a raising operator if $n<0$ and as a lowering operator if $n>0$. 

From every primary field, a tower of states or descendant appears, written
\be
|n_1,n_2,...,;\bar{n}_1,\bar{n}_2,...;\Delta,\bar{\Delta}\ra= \bar{L}_{-\bar{n}_1} \bar{L}_{-\bar{n}_2} ... L_{-n_1} L_{-n_2}... |\Delta,\bar{\Delta}\ra
\ee
and known as Verma module which are eigenstate of $L_0$ with eigenvalue $\Delta+\sum_i n_i$, and the eigenvalue of the antiholomorphic part $\bar{L}_0$ is $\bar{\Delta}+\sum_i \bar{n}_i$. It can happen that not all the descendant are linearly independent, then, it can appear null vectors. So, a final representation of the Virasoro algebra is given by the highest weight and their descendant removing the null vectors. 

The characterization of the null vectors will give us information about the unitarity in a theory. In an infinite representation, a theory is unitary if there is no state with negative norm. For example, we already know that,
\be
\la \Delta|\Delta \ra=1;~~~~\la \Delta| L_1L_{-1}|\Delta\ra=2 \Delta;~~~~\la 0 |L_2L_{-2}|0\ra = \frac{c}{2};
\ee 
then, a necessary condition for unitarity is that $c\ge0$ and $\Delta \ge 0$. Performing the same analysis at the second level in the descendant, there are two independent vectors $\left(L_{-1}\right)^2|\Delta \ra$ and $L_{-2} |\Delta \ra$. Then, it appears a determinant that must be positive defined in a unitary theory,
\be
\begin{vmatrix} \la \Delta | \begin{pmatrix}L_2\\ \left(L_1\right)^2 \end{pmatrix}  \begin{pmatrix}L_{-2},&\left(L_{-1}\right)^2 \end{pmatrix} |\Delta \ra \end{vmatrix}=\begin{vmatrix} 4\Delta +\frac{c}{2} & 6\Delta \\ 6 \Delta & 4\Delta\left( 2\Delta+1\right) \end{vmatrix}
\ee
which is zero iff
\be
\Delta(c)=\frac{5-c\pm \sqrt{(1-c)(25-c)}}{16}
\ee
which shows that theories with $c>25$ are nonunitaries and for $c<1$ it is possible to find theories that are nonunitary. Following in the same way, extracting a determinant at every level, it is possible to obtain the conditions that have to be fulfill a theory to be unitary. These determinants are called Kac determinants and they show that
\begin{enumerate}
\item If $c\ge1$, then unitarity only constraints the values of $\Delta \ge0$
\item If $c<1$, then the set of possible unitary theories is discrete with
\be
c=1-\frac{6}{m(m+1)}; ~~~\Delta_{p,q}(m)=\frac{\left( (m+1)p-mq \right)^2 -1 }{4m(m+1)}; 
\ee
with $1\le q\le p\le m-1$ and for every value of the central charge $c$ there are $\frac{m(m-1)}{2}$ values of the conformal dimensions $\Delta$.
\end{enumerate}
The conditions for unitariy were shown by Friedan, Qiu and Shenker\cite{Friedan:1983xq} and Goddard, Kent and Olive\cite{Goddard:1986ee} proved that they were necessary and sufficient.
\end{document}